\documentclass[11pt]{article}

\usepackage{amsfonts,amssymb,chet,dsfont,graphicx,mathrsfs,pgfplots,tikz}
\allowdisplaybreaks

\newcommand{\1}{\mathds{1}}

\newcommand{\Op}[2]{\mathcal{O}_{#1}(\eta_{#2})}

\newcommand{\ee}[3]{(\eta_{#1}\cdot\eta_{#2})^{#3}}
\newcommand{\e}[3]{\eta_{#1}^{#2_{#3}}}
\newcommand{\D}{\mathcal{D}}
\newcommand{\A}{\mathcal{A}}
\newcommand{\cOPE}[4]{{}_{#1}c_{#2#3}^{\phantom{#2#3}#4}}

\newcommand{\tOPE}[6]{{}_{#1}t_{#2#3}^{#5#6#4}}
\newcommand{\cCF}[4]{{}_{#1}c_{#2#3#4}}
\newcommand{\tCF}[6]{{}_{#1}t_{#2#3#4}^{#5#6}}
\newcommand{\vCF}[6]{{}_{#1}v_{#2#3#4}^{#5#6}}

\newcommand{\Vev}[1]{\left\langle{#1}\right\rangle}

\title{Conformal Three-Point Correlation Functions\\from the Operator Product Expansion}

\author{Jean-Fran\c{c}ois Fortin$^{\ast,}$\email{jean-francois.fortin@phy.ulaval.ca}, Valentina Prilepina$^{\ast,}$\email{valentina.prilepina.1@ulaval.ca} and Witold Skiba$^{\dagger,}$\email{witold.skiba@yale.edu}}

\affiliation{
$^\ast$D\'epartement de Physique, de G\'enie Physique et d'Optique\\Universit\'e Laval, Qu\'ebec, QC G1V 0A6, Canada\\
$^\dagger$Department of Physics, Yale University, New Haven, CT 06520, USA
}

\abstract{We show how to construct embedding space three-point functions for operators in arbitrary Lorentz representations by employing the formalism developed in \cite{Fortin:2019fvx,Fortin:2019dnq}.  We study tensor structures that intertwine the operators with the derivatives in the OPE and examine properties of OPE coefficients under permutations of operators.  Several examples are worked out in detail.  We point out that the group theoretic objects used in this work can be applied directly to construct three-point functions without any reference to the OPE.}

\date{July 2019} 

\begin{document}

\maketitle



\section{Introduction}\label{SecIntro}

Conformal field theories (CFTs) have numerous possible applications to high-energy physics and condensed matter theory and can even be exploited to study gravity via the AdS/CFT correspondence.  They also 
provide a viable avenue for exploring non-conformal theories when such theories are linked by renormalization group flow to CFTs.  The additional symmetries render conformal theories more tractable, in comparison to their relativistic counterparts, allowing CFTs to be formulated non-perturbatively.  The bootstrap approach proposed in the seventies \cite{Ferrara:1973yt,Polyakov:1974gs} exemplifies how the enhanced symmetries of the conformal group constrain CFTs, making non-perturbative treatment possible.

Since correlation functions are the natural observables in CFTs, the ability to effectively compute arbitrary $M$-point functions is essential to studies of such theories.  Two elements are instrumental for the computation of correlation functions.  One such powerful property is the absolute convergence of the operator product expansion (OPE) in CFTs \cite{Mack:1976pa}.  This allows the OPE to be applied successively to sequentially reduce higher-point correlation functions to lower-point ones, ultimately resulting in three- and two-point functions.  The conformal symmetry in $d$ dimensions is linearly realized in the $(d+2)$-dimensional embedding space \cite{Dirac:1936fq}, where the conformal group acts as the Lorentz group, making computations much more transparent.  The application of the OPE in the embedding space was initiated in \cite{Ferrara:1971vh,Ferrara:1971zy,Ferrara:1972cq,Ferrara:1973eg}, while recent advances in \cite{Fortin:2016lmf,Fortin:2016dlj,Comeau:2019xco} led to a novel formulation in \cite{Fortin:2019fvx,Fortin:2019dnq}.

The formalism presented in \cite{Fortin:2019fvx,Fortin:2019dnq} allows one to systematically compute arbitrary $M$-point functions of quasi-primary operators.  In this approach, operators in nontrivial Lorentz representations need to be uplifted to the embedding space, with the chosen embedding exclusively relying on spinor representations, in contrast to, for example, the method in \cite{Mack:1969rr,Weinberg:2012mz}.  The approach uses the fact that any irreducible representation can be obtained from a tensor product of spinor representations.  In this way, all representations are treated in effectively the same manner, as tensors of a single spinor representation in odd dimensions, and tensors of two spinor representations in even dimensions.  The OPE formulated in the embedding space contains both fractional derivatives and ordinary derivatives; however, the action of such derivatives was computed analytically in \cite{Fortin:2019dnq} for any possible expression which may arise in $M$-point functions.

In \cite{Fortin:2019xyr} we applied this new formalism to compute two-point functions of arbitrary quasi-primary operators.  Here, we take the next step and explicitly demonstrate how to obtain three-point functions of operators in arbitrary Lorentz representations.  This work serves both as an illustration and as an explicit test of the methods in \cite{Fortin:2019fvx,Fortin:2019dnq} (various examples of three-point functions are contained in \cite{Rattazzi:2010gj,Poland:2010wg,Costa:2011mg,SimmonsDuffin:2012uy,Hartman:2016dxc,Hofman:2016awc,Hartman:2016lgu,Afkhami-Jeddi:2016ntf,Cuomo:2017wme}).  There are several new components arising in the context of three-point functions, which did not appear in two-point function calculations.  First, there is a tensor structure that intertwines three representations, on both the left- and right-hand sides of the OPE, with the differential operator, which is in the symmetric representation.  Second, the OPE coefficients appear in their most general form here.  Because the embedding space OPE is necessarily not symmetric under operator interchanges, we examine the symmetry properties of these coefficients under operator permutations in three-point functions.  The results obtained here will be useful for a future study of four-point functions.

This paper is organized as follows: in Section \ref{SecThree} we compute three-point correlation functions from the OPE in terms of a specific tensorial quantity.  This tensorial quantity is built from the contraction of hatted projection operators at different embedding space coordinates.  Our next step is to implement a simple conformal substitution rule, which originates from the action of the differential operator appearing in the OPE.  We then cast the result as a tensorial linear combination of simple functions satisfying useful contiguous relations.  The final step is to contract these functions with the appropriate tensor structure to yield the three-point correlation function of interest.  Next, in Section \ref{SecTensor} we discuss the tensor structures in greater detail.  We point out that they form a basis that can always be orthonormalized.  We then describe a general algorithm on how to construct the most general basis of tensor structures for a given three-point correlation function.  In Section \ref{SecSym}, we investigate the symmetry properties of the OPE coefficients by permuting quasi-primary operators in any general three-point correlation function.  Several examples of three-point correlation functions are then presented in Section \ref{SecExamples}.  Some simple cases are projected onto position space to compare with the known results and verify the validity of the formalism introduced in \cite{Fortin:2019fvx,Fortin:2019dnq}.  Subsequently, in Section \ref{SecThreenoOPE} we examine the general form of three-point correlation functions without resorting to the OPE.  Simple arguments are presented on how all contributions can be obtained straightforwardly from a fixed set of objects.  We conclude in Section \ref{SecConc}.

Throughout this paper, we focus on CFTs in Lorentzian signature.  As such, conjugate quasi-primary operators play the role of quasi-primary operators in contragredient-reflected representations.  Our notation follows that of \cite{Fortin:2019dnq,Fortin:2019xyr}.


\section{Three-Point Correlation Functions}\label{SecThree}

In this section, we apply the OPE to write down general expressions for arbitrary three-point correlation functions in embedding space using the new uplift presented in \cite{Fortin:2019fvx,Fortin:2019dnq}.  The result consists of several distinct pieces.  Each external operator is associated with a half-projector that translates a product of quantities carrying spinor indices into a corresponding object in an irreducible Lorentz representation denoted by dummy indices that are contracted with other objects in the three-point function.  The operator that is being exchanged via the OPE gives rise to a group theoretic object whose structure depends exclusively on the Lorentz quantum numbers of the operator along with a Lorentz-singlet part that depends on the dimension of the exchanged operator.  This combination is subsequently transformed into a tensorial function appearing in the three-point function by a straightforward substitution.  Finally, the tensorial function of the coordinates and the dummy indices on the half-projectors are combined into a singlet using a tensor structure.  These tensor structures will be discussed in detail in the next section.


\subsection{The OPE and Three-Point Correlation Functions}

Applying the OPE \cite{Fortin:2019fvx,Fortin:2019dnq}
\eqn{
\begin{gathered}
\Op{i}{1}\Op{j}{2}=(\mathcal{T}_{12}^{\boldsymbol{N}_i}\Gamma)(\mathcal{T}_{21}^{\boldsymbol{N}_j}\Gamma)\cdot\sum_k\sum_{a=1}^{N_{ijk}}\frac{\cOPE{a}{i}{j}{k}\tOPE{a}{i}{j}{k}{1}{2}}{\ee{1}{2}{p_{ijk}}}\cdot\D_{12}^{(d,h_{ijk}-n_a/2,n_a)}(\mathcal{T}_{12\boldsymbol{N}_k}\Gamma)*\Op{k}{2},\\
p_{ijk}=\frac{1}{2}(\tau_i+\tau_j-\tau_k),\qquad h_{ijk}=-\frac{1}{2}(\chi_i-\chi_j+\chi_k),\\
\tau_\mathcal{O}=\Delta_\mathcal{O}-S_\mathcal{O},\qquad\chi_\mathcal{O}=\Delta_\mathcal{O}-\xi_\mathcal{O},\qquad\xi_\mathcal{O}=S_\mathcal{O}-\lfloor S_\mathcal{O}\rfloor,
\end{gathered}
}[EqOPE]
on the first two quasi-primary operators, all three-point correlation functions can be computed with the help of two-point correlation functions \cite{Fortin:2019xyr}, which in turn are also computable from the OPE \eqref{EqOPE}.  Explicitly, all three-point correlation functions are given by
\begingroup\makeatletter\def\f@size{10}\check@mathfonts\def\maketag@@@#1{\hbox{\m@th\large\normalfont#1}}%
\eqna{
\Vev{\Op{i}{1}\Op{j}{2}\Op{k}{3}}&=\frac{(\mathcal{T}_{12}^{\boldsymbol{N}_i}\Gamma)(\mathcal{T}_{21}^{\boldsymbol{N}_j}\Gamma)(\mathcal{T}_{31}^{\boldsymbol{N}_k}\Gamma)}{\ee{1}{2}{\frac{1}{2}(\tau_i+\tau_j-\chi_k)}\ee{1}{3}{\frac{1}{2}(\chi_i-\chi_j+\tau_k)}\ee{2}{3}{\frac{1}{2}(-\chi_i+\chi_j+\chi_k)}}\cdot\sum_{a=1}^{N_{ijk}}\lambda_{\boldsymbol{N}_k}\cCF{a}{i}{j}{k}\,\tCF{a}{i}{j}{k}{1}{2}\\
&\phantom{=}\qquad\cdot\frac{\ee{2}{3}{\chi_k+h_{ijk}}}{\ee{1}{3}{h_{ijk}}}\D_{12}^{(d,h_{ijk}-n_a/2,n_a)}\frac{1}{\ee{2}{3}{\chi_k}}\left(\frac{\eta_3\cdot\Gamma\,\hat{\mathcal{P}}_{32}^{\boldsymbol{N}_k}\cdot\hat{\mathcal{P}}_{12}^{\boldsymbol{N}_k}\,\eta_2\cdot\Gamma}{\ee{2}{3}{}}\right),
}[EqCF]
\endgroup
where the associated three-point function quantities are defined as
\eqn{\cCF{a}{i}{j}{k}=\sum_\ell\cOPE{a}{i}{j}{\ell}\cOPE{}{\ell}{k}{\1},\qquad\tCF{a}{i}{j}{k}{1}{2}=\tOPE{a}{i}{j}{k^C}{1}{2}[(C_\Gamma^{-1})]^{2\xi_k}(g)^{n_v^k}(g)^{n_a},}[EqCoeff]
and $\lambda_{\boldsymbol{N}_k}$ is a normalization constant orthonormalizing the two-point tensor structures \cite{Fortin:2019xyr}.

The structure of \eqref{EqCF} becomes more transparent after explicitly exposing the indices.  The left-hand side of the three-point function has spinor indices for the three quasi-primary operators which are carried by the half-projectors on the right-hand side.  The half-projectors also carry dummy indices, given by $(\mathcal{T}_{12}^{\boldsymbol{N}_i}\Gamma)^{\{Aa\}}$, $(\mathcal{T}_{21}^{\boldsymbol{N}_j}\Gamma)^{\{Bb\}}$, and $(\mathcal{T}_{31}^{\boldsymbol{N}_k}\Gamma)^{\{Cc\}}$, respectively.  These dummy indices are (mostly) contracted with the correlation function tensor structures $(\tCF{a}{i}{j}{k}{1}{2})_{\{aA\}\{bB\}\{c'C'\}\{D\}}$, where the set of indices $\{D\}$ is contracted with the differential operator $\D_{12}^{(d,h_{ijk}-n_a/2,n_a)\{D\}}$.  Hence, the correlation function tensor structures in \eqref{EqCoeff} are obtained from the OPE tensor structures \eqref{EqOPE} by contraction as in
\eqn{(\tCF{a}{i}{j}{k}{1}{2})_{\{aA\}\{bB\}\{c'C'\}\{D\}}=(\tOPE{a}{i}{j}{k^C}{1}{2})_{\{aA\}\{bB\}}^{\phantom{\{aA\}\{bB\}}\{C''c''\}\{D'\}}[(C_\Gamma^{-1})_{c''c'}]^{2\xi_k}(g_{C''C'})^{n_v^k}(g_{D'D})^{n_a}.}[EqTS]
The tensor structures are discussed in more detail in Section \ref{SecTensor}.  To obtain the three-point function, it is still necessary to apply the differential operator in \eqref{EqCF}.  However, as we show in the next section, this is in fact trivial, and the action of the differential operator may be encoded in a simple substitution rule, allowing us to effortlessly obtain any three-point function.


\subsection{Rules for Three-Point Correlation Functions}

Prior to acting with the differential operator, we make several interesting observations about \eqref{EqCF}.  First, the differential operator in \eqref{EqCF} clearly acts on a group theoretic part, represented by
\eqn{\left(\frac{\eta_3\cdot\Gamma\,\hat{\mathcal{P}}_{32}^{\boldsymbol{N}}\cdot\hat{\mathcal{P}}_{12}^{\boldsymbol{N}}\,\eta_2\cdot\Gamma}{\ee{2}{3}{}}\right),}
multiplied by a conformal part, represented by $\ee{2}{3}{-\chi}$, for the exchanged quasi-primary operator.  The group theoretic part encodes all the information about the Lorentz representation of the exchanged quasi-primary operator, while the conformal part depends solely on its conformal dimension.

Second, the last line in \eqref{EqCF} is obviously homogeneous of degree zero in all three embedding space coordinates.  It can thus be expressed in terms of the homogeneized embedding space coordinates, denoted by a bar, which are defined as
\eqn{\bar{\eta}_i^A=\frac{\ee{j}{k}{\frac{1}{2}}}{\ee{i}{j}{\frac{1}{2}}\ee{i}{k}{\frac{1}{2}}}\e{i}{A}{},}[Eqetab]
with $(i,j,k)$ denoting a cyclic permutation of $(1,2,3)$.\footnote{All other permutations are redundant.}

Third, from the general result in \cite{Fortin:2019fvx,Fortin:2019dnq}, the action of the differential operator is simply given by
\eqn{\frac{\ee{2}{3}{p+h+\frac{n-k}{2}}}{\ee{1}{2}{\frac{k}{2}}\ee{1}{3}{h+\frac{n-k}{2}}}\D_{12}^{(d,h,n)}\frac{(\eta_2)^k}{\ee{2}{3}{p}}=\bar{I}_{12}^{(d,h-k,n+k;p)},}[EqD]
where the three-point tensorial function $\bar{I}_{12}^{(d,h,n;p)}$ is described below.

Therefore, the action of the differential operator on the product of the purely group theoretic quantity and the Lorentz-scalar conformal part produces a tensorial object with the proper conformal behavior for three-point correlation functions.  Indeed, acting with the differential operator in the last line of \eqref{EqCF} allows us to define
\eqn{\bar{J}_{12;3}^{(d,h,n,\Delta,\boldsymbol{N})}=\frac{\ee{2}{3}{\chi+h}}{\ee{1}{3}{h}}\D_{12}^{(d,h-n/2,n)}\frac{1}{\ee{2}{3}{\chi}}\left(\frac{\eta_3\cdot\Gamma\,\hat{\mathcal{P}}_{32}^{\boldsymbol{N}}\cdot\hat{\mathcal{P}}_{12}^{\boldsymbol{N}}\,\eta_2\cdot\Gamma}{\ee{2}{3}{}}\right),}[EqJb]
which is homogeneous of degree zero and depends primarily on the exchanged quasi-primary operator.  The $\bar{J}$-function has $n+2n_v$ dummy vector indices, with the set of $2n_v$ indices originating from the hatted projection operators.  Using the identity \eqref{EqD}, \eqref{EqJb} can be very simply expressed as the following substitution for the product of homogeneized embedding space coordinates \eqref{Eqetab},
\eqn{\bar{J}_{12;3}^{(d,h,n,\Delta,\boldsymbol{N})}=\left.\bar{\eta}_3\cdot\Gamma\,\hat{\mathcal{P}}_{32}^{\boldsymbol{N}}\cdot\hat{\mathcal{P}}_{12}^{\boldsymbol{N}}\,\bar{\eta}_2\cdot\Gamma\right|_{\substack{(g)^{s_0}(\bar{\eta}_1)^{s_1}(\bar{\eta}_2)^{s_2}(\bar{\eta}_3)^{s_3}\to(g)^{s_0}(\bar{\eta}_1)^{s_1}(\bar{\eta}_3)^{s_3}\\\times\bar{I}_{12}^{(d,h-n/2-s_2,n+s_2;\chi-s_1/2+s_2/2+s_3/2)}}}.}[EqJbSub]

Applying the conformal substitution rule \eqref{EqJbSub} in the three-point correlation functions \eqref{EqCF} yields
\eqna{
\Vev{\Op{i}{1}\Op{j}{2}\Op{k}{3}}&=\frac{(\mathcal{T}_{12}^{\boldsymbol{N}_i}\Gamma)(\mathcal{T}_{21}^{\boldsymbol{N}_j}\Gamma)(\mathcal{T}_{31}^{\boldsymbol{N}_k}\Gamma)}{\ee{1}{2}{\frac{1}{2}(\tau_i+\tau_j-\chi_k)}\ee{1}{3}{\frac{1}{2}(\chi_i-\chi_j+\tau_k)}\ee{2}{3}{\frac{1}{2}(-\chi_i+\chi_j+\chi_k)}}\\
&\phantom{=}\qquad\cdot\sum_{a=1}^{N_{ijk}}\lambda_{\boldsymbol{N}_k}\cCF{a}{i}{j}{k}\bar{J}_{12;3}^{(d,h_{ijk},n_a,\Delta_k,\boldsymbol{N}_k)}\cdot\tCF{a}{i}{j}{k}{1}{2},
}[EqCFSub]
with the indices suppressed, or
\eqna{
\Vev{\Op{i}{1}\Op{j}{2}\Op{k}{3}}&=\frac{(\mathcal{T}_{12}^{\boldsymbol{N}_i}\Gamma)^{\{Aa\}}(\mathcal{T}_{21}^{\boldsymbol{N}_j}\Gamma)^{\{Bb\}}(\mathcal{T}_{31}^{\boldsymbol{N}_k}\Gamma)^{\{Cc\}}}{\ee{1}{2}{\frac{1}{2}(\tau_i+\tau_j-\chi_k)}\ee{1}{3}{\frac{1}{2}(\chi_i-\chi_j+\tau_k)}\ee{2}{3}{\frac{1}{2}(-\chi_i+\chi_j+\chi_k)}}\\
&\phantom{=}\qquad\times\sum_{a=1}^{N_{ijk}}\lambda_{\boldsymbol{N}_k}\cCF{a}{i}{j}{k}\,(\tCF{a}{i}{j}{k}{1}{2})_{\{aA\}\{bB\}\{c'C'\}\{D\}}(\bar{J}_{12;3}^{(d,h_{ijk},n_a,\Delta_k,\boldsymbol{N}_k)})_{\{cC\}}^{\phantom{\{cC\}}\{C'c'\}\{D\}},
}
with the dummy indices exposed.  Equation \eqref{EqCFSub} is valid for all three-point correlation functions with quasi-primary operators in general irreducible representations of the Lorentz group.  The nontrivial part of the conformal substitution rule \eqref{EqJbSub} depends only on the irreducible representation of the exchanged quasi-primary operator.  Although the contractions of the hatted projection operators are not necessarily trivial, the remaining substitution rule is straightforward.

Since the $\bar{J}$-function is obtained from contractions of the hatted projection operator for the representation $\boldsymbol{N}_k$, it is clear that it depends solely on the irreducible representation of the exchanged quasi-primary operator.\footnote{The hatted projection operators are investigated in more detail in \cite{Fortin:2019xyr}.}  Therefore, once the irreducible representation is fixed, the associated $\bar{J}$-function is valid for all quasi-primary operators in this same irreducible representation.  The tensor structures are then responsible for the proper contractions of the remaining dummy indices, as in $\bar{J}_{12;3}^{(d,h_{ijk},n_a,\Delta_k,\boldsymbol{N}_k)}\cdot\tCF{a}{i}{j}{k}{1}{2}$.


\subsection{Three-Point Tensorial Function}

The three-point tensorial function appearing in \eqref{EqD} and the substitution rule \eqref{EqJbSub} is given by \cite{Fortin:2019fvx,Fortin:2019dnq}
\eqn{\bar{I}_{12}^{(d,h,n;p)}=(-2)^h(p)_h(p+1-d/2)_h\sum_{\substack{q_0,q_1,q_2,q_3\geq0\\\bar{q}=2q_0+q_1+q_2+q_3=n}}S_{(q_0,q_1,q_2,q_3)}K^{(d,h;p;q_0,q_1,q_2,q_3)},}[EqIb]
where the fully-symmetric tensor and the $K$-function are
\eqna{
S_{(q_0,q_1,q_2,q_3)}^{A_1\cdots A_{\bar{q}}}&=g^{(A_1A_2}\cdots g^{A_{2q_0-1}A_{2q_0}}\bar{\eta}_1^{A_{2q_0+1}}\cdots\bar{\eta}_1^{A_{2q_0+q_1}}\\
&\phantom{=}\qquad\times\bar{\eta}_2^{A_{2q_0+q_1+1}}\cdots\bar{\eta}_2^{A_{2q_0+q_1+q_2}}\bar{\eta}_3^{A_{2q_0+q_1+q_2+1}}\cdots\bar{\eta}_3^{A_{\bar{q}})},\\
K^{(d,h;p;q_0,q_1,q_2,q_3)}&=\frac{(-1)^{\bar{q}-q_0-q_1-q_2}(-2)^{\bar{q}-q_0}\bar{q}!}{q_0!q_1!q_2!q_3!}\frac{(-h-\bar{q})_{\bar{q}-q_0-q_2}(p+h)_{\bar{q}-q_0-q_1}}{(p+1-d/2)_{-q_0-q_1-q_2}}.
}[EqK]
Here, $\bar{q}=2q_0+q_1+q_2+q_3$.  The three-point tensorial function is totally symmetric and traceless with respect to the embedding space metric.  Moreover, it satisfies several contiguous relations \cite{Fortin:2019fvx,Fortin:2019dnq}, given by
\eqna{
g\cdot\bar{I}_{12}^{(d,h,n;p)}&=0,\\
\bar{\eta}_1\cdot\bar{I}_{12}^{(d,h,n;p)}&=\bar{I}_{12}^{(d,h+1,n-1;p)},\\
\bar{\eta}_2\cdot\bar{I}_{12}^{(d,h,n;p)}&=(-2)(-h-n)(-h-n+1-d/2)\bar{I}_{12}^{(d,h,n-1;p)},\\
\bar{\eta}_3\cdot\bar{I}_{12}^{(d,h,n;p)}&=\bar{I}_{12}^{(d,h+1,n-1;p-1)}.
}[EqCont]
These contiguous relations are very useful in evaluating contractions between the $\bar{J}$-functions and the tensor structures, $\bar{J}_{12;3}^{(d,h_{ijk},n_a,\Delta_k,\boldsymbol{N}_k)}\cdot\tCF{a}{i}{j}{k}{1}{2}$, appearing in the three-point functions \eqref{EqCFSub}.


\section{Tensor Structures}\label{SecTensor}

The correlation function tensor structures \eqref{EqTS} intertwine three irreducible representations of the Lorentz group into a symmetric-traceless irreducible representation.  Indeed, the number of symmetric-traceless irreducible representations in the product $\boldsymbol{N}_i\otimes\boldsymbol{N}_j\otimes\boldsymbol{N}_k$ corresponds to the number $N_{ijk}$ of tensor structures and OPE coefficients appearing in the OPE \eqref{EqOPE}.  In this section, we discuss properties of the tensor structures and present a general algorithm for their construction.


\subsection{Tensor Structure Properties}

Since tensor structures contract four irreducible representations of the Lorentz group together into a singlet, they should be appropriately related to their position space counterparts.  The relationship between the tensor structures in embedding space $\tCF{a}{i}{j}{k}{1}{2}$ and those in position space can be expressed via the simple substitutions \cite{Fortin:2019dnq}
\eqn{
\begin{gathered}
g^{\mu\nu}\to\A_{12}^{AB}=g^{AB}-\frac{\e{1}{A}{}\e{2}{B}{}}{\ee{1}{2}{}}-\frac{\e{1}{B}{}\e{2}{A}{}}{\ee{1}{2}{}},\\
\epsilon^{\mu_1\cdots\mu_d}\to\epsilon_{12}^{A_1\cdots A_d}=\frac{1}{\ee{1}{2}{}}\eta_{1A_0'}\epsilon^{A_0'A_1'\cdots A_d'A_{d+1}'}\eta_{2A_{d+1}'}\A_{12A_d'}^{\phantom{12A_d'}A_d}\cdots\A_{12A_1'}^{\phantom{12A_1'}A_1},\\
\gamma^{\mu_1\cdots\mu_n}\to\Gamma_{12}^{A_1\cdots A_n}=\Gamma^{A_1'\cdots A_n'}\A_{12A_n'}^{\phantom{12A_n'}A_n}\cdots\A_{12A_1'}^{\phantom{12A_1'}A_1}\qquad\forall\,n\in\{0,\ldots,r\}.
\end{gathered}
}[EqTSsub]

From the form of the OPE \eqref{EqOPE}, the correlation function tensor structures must also satisfy the following identity \cite{Fortin:2019dnq}
\eqn{\tCF{a}{i}{j}{k}{1}{2}=(\hat{\mathcal{P}}_{12}^{\boldsymbol{N}_i})(\hat{\mathcal{P}}_{21}^{\boldsymbol{N}_j})(\hat{\mathcal{P}}_{12}^{\boldsymbol{N}_k})(\hat{\mathcal{P}}_{21}^{n_a\boldsymbol{e}_1})\cdot\tCF{a}{i}{j}{k}{1}{2},}[EqPTS]
where the order of the contractions is clear.  Since each set of indices corresponds to particular irreducible representations of the Lorentz group, \eqref{EqPTS} simply forces the correlation function tensor structures to project onto the appropriate irreducible representations for the three quasi-primary operators and the symmetric-traceless differential operator.

Moreover, the set of all tensor structures for a fixed choice of $\boldsymbol{N}_i$, $\boldsymbol{N}_j$ and $\boldsymbol{N}_k$ forms a basis for a vector space, and the structures can be orthonormalized \cite{Fortin:2019dnq}.  Thus, they can be chosen to satisfy
\eqn{\tCF{a}{i}{j}{k}{1}{2}\cdot\tCF{b}{i}{j}{k}{2}{1*}=\delta_{ab},}[EqOrthonorm]
where the normalization constant differs by a factor of two between position space and embedding space tensor structures due to the different sizes of the gamma matrices \cite{Fortin:2019xyr}.  The inner product \eqref{EqOrthonorm} is defined as
\eqna{
\tCF{a}{i}{j}{k}{1}{2}\cdot\tCF{b}{i}{j}{k}{2}{1*}&\equiv(\tCF{a}{i}{j}{k}{1}{2})_{\{aA\}\{bB\}\{cC\}\{D\}}(B_\Gamma^{-1}\tCF{b}{i}{j}{k}{2}{1*}C_\Gamma^*B_\Gamma C_\Gamma^{-1})_{\{a'A'\}\{b'B'\}\{c'C'\}\{D'\}}\\
&\phantom{=}\qquad\times(g^{AA'})^{n_v^i}[(C_\Gamma)^{aa'}]^{2\xi_i}(g^{BB'})^{n_v^j}[(C_\Gamma)^{bb'}]^{2\xi_j}(g^{CC'})^{n_v^k}[(C_\Gamma)^{cc'}]^{2\xi_k}(g^{DD'})^{n_a},
}
where the indices $a$ and $b$ that appear as subscripts in the different tensor structures have nothing to do with the possible embedding spinor indices, coincidentally also denoted as $a$, $b$, $c$ on the right-hand side (hopefully not causing any confusion).  The corresponding vector space is also graded, with the grading associated with the number of tensor structures projecting to the same symmetric-traceless irreducible representation.  This is straightforward, since tensor structures projecting onto different symmetric-traceless irreducible representations are automatically orthogonal.

Dropping all fixed indices on the tensor structures for notational simplicity, it is clear that once an orthonormal basis \eqref{EqOrthonorm} has been found, any unitary transformation ${}_at'={}_a(Ut)\equiv U_{aa'}\,{}_{a'}t$  also forms an orthonormal basis.  Under such a unitary transformation, the correlation function coefficients \eqref{EqCoeff} transform as ${}_ac'={}_a(cU^\dagger)\equiv{}_{a'}c\,U_{aa'}^*$.


\subsection{Algorithm}

Although it might be convenient, it is not necessary to use an orthonormal basis of tensor structures.  Starting from an arbitrary basis of correlation function tensor structures ${}_av$ with inner product ${}_av\cdot{}_bv^*=M_{ab}$, it is always possible to generate an orthonormal basis ${}_at$ using a similarity transformation $S$, such that ${}_at={}_a(Sv)\equiv S_{aa'}\,{}_{a'}v$.  Indeed, the necessary similarity transformation must satisfy
\eqn{\delta_{ab}={}_at\cdot{}_bt^*={}_a(Sv)\cdot{}_b(Sv)^*=(SMS^\dagger)_{ab},}
or $(S^\dagger S)^{-1}=M$.  Thus, $S$ can be decomposed as $S=D^{-1/2}U$, where $U$ is the unitary matrix diagonalizing the Hermitian matrix $M$ of inner products, and $D$ is the diagonal matrix of eigenvalues of $M$.  Therefore, it is always possible to start from an arbitrary basis of correlation function tensor structures ${}_av$ and then orthonormalize them as in \eqref{EqOrthonorm} if so desired.

Since the tensor structures take four irreducible representations of the Lorentz group and combine them into singlets, they must contract all dummy indices of the three quasi-primary operators as well as the indices of the symmetric-traceless differential operator.  Moreover, the tensor structures satisfy \eqref{EqPTS} by construction; hence, they cannot contract dummy indices from the same set (for example, two dummy indices in $\{aA\}$), due to the tracelessness property of the hatted projection operators.  In consequence, it is only necessary to generate all possible contractions of the dummy indices between distinct sets using \eqref{EqTSsub} to generate an arbitrary basis.

In fact, it is also possible to generate a basis without imposing \eqref{EqPTS} first.  In particular, one may simply enumerate all possible allowed independent trace-free contractions.  Such an arbitrary basis might end up being overcomplete, but applying the four hatted projection operators as in \eqref{EqPTS} on it would subsequently reveal that the redundant tensor structures are in fact equivalent and can therefore be discarded.  This results in an nonredundant arbitrary basis, and the procedure highlighted above can be implemented quite easily to arrive at an orthonormal basis.  Several examples will be presented in a subsequent section.


\section{Symmetry Properties of the Correlation Function OPE Coefficients}\label{SecSym}

From \eqref{EqCFSub}, three-point correlation functions are given by
\eqna{
\Vev{\Op{i}{1}\Op{j}{2}\Op{k}{3}}&=\frac{(\mathcal{T}_{12}^{\boldsymbol{N}_i}\Gamma)(\mathcal{T}_{21}^{\boldsymbol{N}_j}\Gamma)(\mathcal{T}_{31}^{\boldsymbol{N}_k}\Gamma)}{\ee{1}{2}{\frac{1}{2}(\tau_i+\tau_j-\chi_k)}\ee{1}{3}{\frac{1}{2}(\chi_i-\chi_j+\tau_k)}\ee{2}{3}{\frac{1}{2}(-\chi_i+\chi_j+\chi_k)}}\\
&\phantom{=}\qquad\cdot\sum_{a=1}^{N_{ijk}}\lambda_{\boldsymbol{N}_k}\cCF{a}{i}{j}{k}\bar{J}_{12;3}^{(d,h_{ijk},n_a,\Delta_k,\boldsymbol{N}_k)}\cdot\tCF{a}{i}{j}{k}{1}{2}.
}[EqCFijk]
Since the OPE is not symmetric by construction, it is of interest to investigate the effect of interchanging the order of the quasi-primary operators.  Permuting operators in \eqref{EqCFijk} results in constraints on the coefficients $\cCF{a}{i}{j}{k}$ under index permutations.  Such constraints are the subject of this section.


\subsection{Permutation of \texorpdfstring{$i$}{i} and \texorpdfstring{$j$}{j}}

Under the exchange $i\leftrightarrow j$, the three-point correlation function becomes
\eqna{
\Vev{\Op{j}{1}\Op{i}{2}\Op{k}{3}}&=\frac{(\mathcal{T}_{12}^{\boldsymbol{N}_j}\Gamma)(\mathcal{T}_{21}^{\boldsymbol{N}_i}\Gamma)(\mathcal{T}_{31}^{\boldsymbol{N}_k}\Gamma)}{\ee{1}{2}{\frac{1}{2}(\tau_j+\tau_i-\chi_k)}\ee{1}{3}{\frac{1}{2}(\chi_j-\chi_i+\tau_k)}\ee{2}{3}{\frac{1}{2}(-\chi_j+\chi_i+\chi_k)}}\\
&\phantom{=}\qquad\cdot\sum_{a=1}^{N_{ijk}}\lambda_{\boldsymbol{N}_k}\cCF{a}{j}{i}{k}\bar{J}_{12;3}^{(d,h_{jik},n_a,\Delta_k,\boldsymbol{N}_k)}\cdot\tCF{a}{j}{i}{k}{1}{2}.
}
Further, interchanging $\eta_1\leftrightarrow\eta_2$ leads to
\eqna{
\Vev{\Op{j}{2}\Op{i}{1}\Op{k}{3}}&=\frac{(\mathcal{T}_{21}^{\boldsymbol{N}_j}\Gamma)(\mathcal{T}_{12}^{\boldsymbol{N}_i}\Gamma)(\mathcal{T}_{32}^{\boldsymbol{N}_k}\Gamma)}{\ee{1}{2}{\frac{1}{2}(\tau_j+\tau_i-\chi_k)}\ee{2}{3}{\frac{1}{2}(\chi_j-\chi_i+\tau_k)}\ee{1}{3}{\frac{1}{2}(-\chi_j+\chi_i+\chi_k)}}\\
&\phantom{=}\qquad\cdot\sum_{a=1}^{N_{ijk}}\lambda_{\boldsymbol{N}_k}\cCF{a}{j}{i}{k}\bar{J}_{21;3}^{(d,h_{jik},n_a,\Delta_k,\boldsymbol{N}_k)}\cdot\tCF{a}{j}{i}{k}{2}{1},
}
which can be directly compared with \eqref{EqCF}.  Indeed, one has
\eqna{
\Vev{\Op{i}{1}\Op{j}{2}\Op{k}{3}}&=(-1)^{\xi_i+\xi_j-\xi_k}\Vev{\Op{j}{2}\Op{i}{1}\Op{k}{3}}\\
&=\frac{(-1)^{\xi_i+\xi_j-\xi_k}(\mathcal{T}_{12}^{\boldsymbol{N}_i}\Gamma)(\mathcal{T}_{21}^{\boldsymbol{N}_j}\Gamma)\left(\mathcal{T}_{31}^{\boldsymbol{N}_k}\Gamma\cdot\frac{\eta_3\cdot\Gamma\,\hat{\mathcal{P}}_{32}^{\boldsymbol{N}_k}\,\eta_2\cdot\Gamma}{2\ee{2}{3}{}}\right)}{\ee{1}{2}{\frac{1}{2}(\tau_i+\tau_j-\chi_k)}\ee{1}{3}{\frac{1}{2}(\chi_i-\chi_j+\tau_k)}\ee{2}{3}{\frac{1}{2}(-\chi_i+\chi_j+\chi_k)}}\\
&\phantom{=}\qquad\cdot\sum_{a=1}^{N_{ijk}}\lambda_{\boldsymbol{N}_k}\cCF{a}{j}{i}{k}\bar{J}_{21;3}^{(d,h_{jik},n_a,\Delta_k,\boldsymbol{N}_k)}\cdot\tCF{a}{j}{i}{k}{2}{1},
}
which corresponds to
\begingroup\makeatletter\def\f@size{10}\check@mathfonts\def\maketag@@@#1{\hbox{\m@th\large\normalfont#1}}%
\eqn{\sum_{a=1}^{N_{ijk}}(-1)^{\xi_i+\xi_j-\xi_k}\lambda_{\boldsymbol{N}_k}\cCF{a}{j}{i}{k}\frac{\eta_3\cdot\Gamma\,\hat{\mathcal{P}}_{32}^{\boldsymbol{N}_k}\,\eta_2\cdot\Gamma}{2^{2\xi_k}\ee{2}{3}{}}\cdot\bar{J}_{21;3}^{(d,h_{jik},n_a,\Delta_k,\boldsymbol{N}_k)}\cdot\tCF{a}{j}{i}{k}{2}{1}=\sum_{a=1}^{N_{ijk}}\lambda_{\boldsymbol{N}_k}\cCF{a}{i}{j}{k}\bar{J}_{12;3}^{(d,h_{ijk},n_a,\Delta_k,\boldsymbol{N}_k)}\cdot\tCF{a}{i}{j}{k}{1}{2}.}[EqPij]
\endgroup
The dummy indices corresponding to the same quasi-primary operators must obviously be chosen to have the same labels on both sides.  Moreover, in \eqref{EqPij} it is not necessarily possible to remove the sums because the tensor structures might not match one to one.  Examples in the next section will make this point clear.  Finally, the factor of $2$ is absent for bosonic representations.


\subsection{Arbitrary Permutations}

Proceeding as before for an arbitrary permutation, the set of symmetry properties of the correlation function OPE coefficients is given by
\eqna{
&\sum_{a=1}^{N_{ijk}}\lambda_{\boldsymbol{N}_k}\cCF{a}{i}{j}{k}\bar{J}_{12;3}^{(d,h_{ijk},n_a,\Delta_k,\boldsymbol{N}_k)}\cdot\tCF{a}{i}{j}{k}{1}{2}\\
&\qquad=\sum_{a=1}^{N_{ijk}}(-1)^{\xi_i+\xi_j-\xi_k}\lambda_{\boldsymbol{N}_k}\cCF{a}{j}{i}{k}\frac{\eta_3\cdot\Gamma\,\hat{\mathcal{P}}_{32}^{\boldsymbol{N}_k}\,\eta_2\cdot\Gamma}{2^{2\xi_k}\ee{2}{3}{}}\cdot\bar{J}_{21;3}^{(d,h_{jik},n_a,\Delta_k,\boldsymbol{N}_k)}\cdot\tCF{a}{j}{i}{k}{2}{1}\\
&\qquad=\sum_{a=1}^{N_{ijk}}(-1)^{\xi_i+\xi_j+\xi_k}\lambda_{\boldsymbol{N}_i}\cCF{a}{k}{j}{i}\frac{\eta_1\cdot\Gamma\,\hat{\mathcal{P}}_{13}^{\boldsymbol{N}_i}\,\eta_3\cdot\Gamma}{2^{2\xi_i}\ee{1}{3}{}}\cdot\bar{J}_{32;1}^{(d,h_{kji},n_a,\Delta_i,\boldsymbol{N}_i)}\\
&\qquad\phantom{=}\qquad\cdot\left(\frac{\eta_3\cdot\Gamma\,\hat{\mathcal{P}}_{32}^{\boldsymbol{N}_k}\,\eta_2\cdot\Gamma}{2^{2\xi_k}\ee{2}{3}{}}\frac{\eta_2\cdot\Gamma\,\hat{\mathcal{P}}_{23}^{\boldsymbol{N}_j}\,\eta_3\cdot\Gamma}{2^{2\xi_j}\ee{2}{3}{}}\cdot\tCF{a}{k}{j}{i}{3}{2}\right)\\
&\qquad=\sum_{a=1}^{N_{ijk}}(-1)^{-\xi_i+\xi_j+\xi_k}\lambda_{\boldsymbol{N}_j}\cCF{a}{i}{k}{j}\bar{J}_{13;2}^{(d,h_{ikj},n_a,\Delta_j,\boldsymbol{N}_j)}\cdot\left(\frac{\eta_1\cdot\Gamma\,\hat{\mathcal{P}}_{13}^{\boldsymbol{N}_i}\,\eta_3\cdot\Gamma}{2^{2\xi_i}\ee{1}{3}{}}\cdot\tCF{a}{i}{k}{j}{1}{3}\right)\\
&\qquad=\sum_{a=1}^{N_{ijk}}(-1)^{2\xi_i}\lambda_{\boldsymbol{N}_i}\cCF{a}{j}{k}{i}\bar{J}_{23;1}^{(d,h_{jki},n_a,\Delta_i,\boldsymbol{N}_i)}\cdot\left(\frac{\eta_2\cdot\Gamma\,\hat{\mathcal{P}}_{23}^{\boldsymbol{N}_j}\,\eta_3\cdot\Gamma}{2^{2\xi_j}\ee{2}{3}{}}\frac{\eta_3\cdot\Gamma\,\hat{\mathcal{P}}_{32}^{\boldsymbol{N}_k}\,\eta_2\cdot\Gamma}{2^{2\xi_k}\ee{2}{3}{}}\cdot\tCF{a}{j}{k}{i}{2}{3}\right)\\
&\qquad=\sum_{a=1}^{N_{ijk}}(-1)^{2\xi_k}\lambda_{\boldsymbol{N}_j}\cCF{a}{k}{i}{j}\frac{\eta_2\cdot\Gamma\,\hat{\mathcal{P}}_{23}^{\boldsymbol{N}_j}\,\eta_3\cdot\Gamma}{2^{2\xi_j}\ee{2}{3}{}}\cdot\bar{J}_{31;2}^{(d,h_{kij},n_a,\Delta_j,\boldsymbol{N}_j)}\cdot\left(\frac{\eta_1\cdot\Gamma\,\hat{\mathcal{P}}_{13}^{\boldsymbol{N}_i}\,\eta_3\cdot\Gamma}{2^{2\xi_i}\ee{1}{3}{}}\cdot\tCF{a}{k}{i}{j}{3}{1}\right).
}
\eqn{}[EqPerm]
In \eqref{EqPerm}, the proper contractions should be obvious from the context.  Again, the factors of $2$ are absent for bosonic representations.

The identities \eqref{EqPerm} relate all permutations of the correlation function OPE coefficients to each other.  Although seemingly complicated, they simply represent the equality of the original three-point correlation function under the choice of quasi-primary operators on which the OPE \eqref{EqOPE} is used.  The specific examples presented below will clarify these identities.


\section{Examples of Three-Point Correlation Functions}\label{SecExamples}

In this section, it is shown how to compute the $\bar{J}$-functions for specific explicit irreducible representations of the Lorentz group.  Furthermore, we construct the tensor structures for some three-point correlation functions of interest.  These depend on all three quasi-primary operators appearing in the given three-point function.  To  illustrate the approach, the contractions of the dummy indices between the $\bar{J}$-function and the correlation function tensor structures are carried out in detail, with the contiguous relations \eqref{EqCont} employed to simplify the results, in order to explicitly compute three-point functions in a few cases.  In several examples, projection to position space is presented to confirm that the method works properly.

It is important to emphasize that this formalism holds for any irreducible representation.  Hence, once the hatted projection operator for a given representation is known, it is straightforward to then apply the substitution rule \eqref{EqJbSub} to compute the corresponding $\bar{J}$-function.


\subsection{\texorpdfstring{$\bar{J}$}{Jb}-function}

As previously mentioned, the $\bar{J}$-function \eqref{EqJbSub} can be rewritten to simplify the notation as
\eqna{
\bar{J}_{12;3}^{(d,h,n,\Delta,\boldsymbol{N})}&=(\bar{\eta}_3\cdot\Gamma\,\hat{\mathcal{P}}_{32}^{\boldsymbol{N}}\cdot\hat{\mathcal{P}}_{12}^{\boldsymbol{N}}\,\bar{\eta}_2\cdot\Gamma)_{cs}\\
&\equiv\left.\bar{\eta}_3\cdot\Gamma\,\hat{\mathcal{P}}_{32}^{\boldsymbol{N}}\cdot\hat{\mathcal{P}}_{12}^{\boldsymbol{N}}\,\bar{\eta}_2\cdot\Gamma\right|_{\substack{(g)^{s_0}(\bar{\eta}_1)^{s_1}(\bar{\eta}_2)^{s_2}(\bar{\eta}_3)^{s_3}\to(g)^{s_0}(\bar{\eta}_1)^{s_1}(\bar{\eta}_3)^{s_3}\\\times\bar{I}_{12}^{(d,h-n/2-s_2,n+s_2;\chi-s_1/2+s_2/2+s_3/2)}}},
}[EqJbSubs]
where the subscript $cs$ denotes the conformal substitution rule \eqref{EqJbSub}.  This function depends primarily on the irreducible representation of the exchanged quasi-primary operator.  To compute $\bar{J}$, one needs to contract two hatted projection operators at different embedding space coordinates.  Clearly, the result would be trivial for two projection operators at the same embedding space coordinates, by definition.  However, in \eqref{EqJbSubs} there are three different embedding space points.  Nevertheless, because the hatted projection operators share a coordinate and because $\A_{ij}\cdot\A_{jk}\cdot\A_{ij}=\A_{ij}$, the trace parts of one of the two hatted projection operators vanish identically, somewhat simplifying the final result.

Although \eqref{EqJbSubs} seems complicated for large irreducible representations, it is in fact a trivial matter to implement the substitution rule \eqref{EqJbSub} in any symbolic manipulation program.


\subsubsection{Symmetric-Traceless Exchange}

We first turn to the case of symmetric-traceless exchange.  For quasi-primary operators in the symmetric-traceless irreducible representation $\ell\boldsymbol{e}_1$, the hatted projection operator is given by
\eqna{
(\hat{\mathcal{P}}_{12}^{\ell\boldsymbol{e}_1})_{A_\ell\cdots A_1}^{\phantom{A_\ell\cdots A_1}B_1\cdots B_\ell}&=\sum_{i=0}^{\lfloor\ell/2\rfloor}\frac{(-\ell)_{2i}}{2^{2i}i!(-\ell+2-d/2)_i}\A_{12(A_1A_2}\A_{12}^{(B_1B_2}\cdots\A_{12A_{2i-1}A_{2i}}\A_{12}^{B_{2i-1}B_{2i}}\\
&\phantom{=}\qquad\times\A_{12A_{2i+1}}^{\phantom{12A_{2i+1}}B_{2i+1}}\cdots\A_{12A_\ell)}^{\phantom{12A_\ell)}B_\ell)}.
}[EqPle1]
Therefore, the $\bar{J}$-function \eqref{EqJbSubs} becomes
\eqna{
(\bar{J}_{12;3}^{(d,h,n,\Delta,\ell\boldsymbol{e}_1)})_{\{C\}}^{\phantom{\{C\}}\{C'\}\{D\}}&=\sum_{i=0}^{\lfloor\ell/2\rfloor}\frac{(-\ell)_{2i}}{2^{2i}i!(-\ell+2-d/2)_i}\left(\A_{23(C_1C_2}\A_{12}^{(C_1'C_2'}\cdots\A_{23C_{2i-1}C_{2i}}\A_{12}^{C_{2i-1}'C_{2i}'}\right.\\
&\phantom{=}\qquad\times\left.\A_{321C_{2i+1}}^{\phantom{321C_{2i+1}}C_{2i+1}'}\cdots\A_{321C_\ell)}^{\phantom{321C_\ell)}C_\ell')}\right)_{cs},
}[EqJble1]
where $\A_{ijk}=\A_{ij}\cdot\A_{jk}$.  This result clearly illustrates the point about the vanishing of trace terms, as the double sum from two $\hat{\mathcal{P}}^{\ell\boldsymbol{e}_1}$'s collapses into a single sum.  The set of dummy indices $\{D\}$ seemingly missing on the right-hand side of \eqref{EqJble1} is implicitly included in the definition of the tensorial function \eqref{EqIb}, appearing once the substitution rule is used.  Given \eqref{EqJble1}, it is a simple matter to compute any three-point correlation function with an exchange of quasi-primary operators in symmetric-traceless representations.

The $\bar{J}$-functions for quasi-primary operators in irreducible representations $\boldsymbol{0}$ and $\boldsymbol{e}_1$, respectively, are explicitly given by
\eqna{
(\bar{J}_{12;3}^{(d,h,n,\Delta,\boldsymbol{0})})^{\{D\}}&=(1)_{cs}=\bar{I}_{12}^{(d,h-n/2,n;\Delta)\{D\}},\\
(\bar{J}_{12;3}^{(d,h,n,\Delta,\boldsymbol{e}_1)})_C^{\phantom{C}C'\{D\}}&=(\A_{321C}^{\phantom{321C}C'})_{cs}=(g_C^{\phantom{C}C'}-\bar{\eta}_{1C}\bar{\eta}_2^{C'}-\bar{\eta}_{2C}\bar{\eta}_3^{C'}+\bar{\eta}_{2C}\bar{\eta}_2^{C'})_{cs}\\
&=g_C^{\phantom{C}C'}\bar{I}_{12}^{(d,h-n/2,n;\Delta)\{D\}}-\bar{\eta}_{1C}\bar{I}_{12}^{(d,h-n/2-1,n+1;\Delta)C'\{D\}}\\
&\phantom{=}\qquad-\bar{\eta}_3^{C'}\bar{I}_{12}^{(d,h-n/2-1,n+1;\Delta+1)}{}_C^{\phantom{C}\{D\}}+\bar{I}_{12}^{(d,h-n/2-2,n+2;\Delta+1)}{}_C^{\phantom{C}C'\{D\}},
}[EqJb0ande1]
once the conformal substitution rule \eqref{EqJbSub} has been applied.


\subsubsection{Spinor Exchange}

Proceeding to the case of spinor exchange, we note that the hatted projection operator for quasi-primary operators in the spinor irreducible representation $\boldsymbol{e}_r$ is given by
\eqn{(\hat{\mathcal{P}}_{12}^{\boldsymbol{e}_r})_a^{\phantom{a}b}=\delta_a^{\phantom{a}b},}[EqPer]
assuming odd $d$ for simplicity.  Hence, the precise form of the $\bar{J}$-function \eqref{EqJbSubs} is simply
\eqn{(\bar{J}_{12;3}^{(d,h,n,\Delta,\boldsymbol{e}_r)})_c^{\phantom{c}c'\{D\}}=[(\bar{\eta}_3\cdot\Gamma\bar{\eta}_2\cdot\Gamma)_c^{\phantom{c}c'}]_{cs}=(\bar{\eta}_3\cdot\Gamma\Gamma_{D_0})_c^{\phantom{c}c'}\bar{I}_{12}^{(d,h-n/2-1,n+1;\Delta+1/2)D_0\{D\}},}[EqJber]
which can be used in all three-point correlation functions where quasi-primary operators in spinor irreducible representations are exchanged.


\subsubsection{\texorpdfstring{$\boldsymbol{e}_1+\boldsymbol{e}_r$}{e1+er} Exchange}

To demonstrate the universality of the method, it is of interest to determine the $\bar{J}$-function for quasi-primary operators in some mixed irreducible representation of the Lorentz group.  Again assuming odd $d$ for simplicity, the $\bar{J}$-function \eqref{EqJbSubs} for quasi-primary operators in the irreducible representation $\boldsymbol{e}_1+\boldsymbol{e}_r$ is given by
\eqna{
(\bar{J}_{12;3}^{(d,h,n,\Delta,\boldsymbol{e}_1+\boldsymbol{e}_r)})_{cC}^{\phantom{cC}C'c'}&=\left[\A_{321C}^{\phantom{321C}C'}(\bar{\eta}_3\cdot\Gamma\bar{\eta}_2\cdot\Gamma)_c^{\phantom{c}c'}-\frac{1}{d}(\bar{\eta}_3\cdot\Gamma\Gamma_{32C}\Gamma_{12}^{C'}\bar{\eta}_2\cdot\Gamma)_c^{\phantom{c}c'}\right]_{cs}\\
&=(\bar{\eta}_3\cdot\Gamma\Gamma_{D_0})_c^{\phantom{c}c'}\left[g_C^{\phantom{C}C'}\bar{I}_{12}^{(d,h-n/2-1,n+1;\Delta+1)D_0\{D\}}\right.\\
&\phantom{=}\left.\qquad-\bar{\eta}_{1C}\bar{I}_{12}^{(d,h-n/2-2,n+2;\Delta+1)C'D_0\{D\}}\right.\\
&\phantom{=}\left.\qquad-\bar{\eta}_3^{C'}\bar{I}_{12}^{(d,h-n/2-2,n+2;\Delta+2)}{}_C^{\phantom{C}D_0\{D\}}+\bar{I}_{12}^{(d,h-n/2-3,n+3;\Delta+2)}{}_C^{\phantom{C}C'D_0\{D\}}\right]\\
&\phantom{=}\qquad-\frac{1}{d}(\bar{\eta}_3\cdot\Gamma\Gamma_C\Gamma^{C'}\Gamma_{D_0})_c^{\phantom{c}c'}\bar{I}_{12}^{(d,h-n/2-1,n+1;\Delta+1)D_0\{D\}}\\
&\phantom{=}\qquad+\frac{1}{d}(\bar{\eta}_3\cdot\Gamma\Gamma_C\bar{\eta}_1\cdot\Gamma\Gamma_{D_0})_c^{\phantom{c}c'}\bar{I}_{12}^{(d,h-n/2-2,n+2;\Delta+1)C'D_0\{D\}},
}[EqJbe1+er]
since the hatted projection operator is simply
\eqn{(\hat{\mathcal{P}}_{12}^{\boldsymbol{e}_1+\boldsymbol{e}_r})_{aA}^{\phantom{aA}Bb}=\A_{12A}^{\phantom{12A}B}\delta_a^{\phantom{a}b}-\frac{1}{d}(\Gamma_{12A}\Gamma_{12}^B)_a^{\phantom{a}b}.}[EqPe1+er]
From the first equality in \eqref{EqJbe1+er}, it is easy to see the vanishing of trace terms also for fermionic representations.

In order to compute explicit three-point correlation functions with exchanged quasi-primary operators in the irreducible representations shown above, it necessary to determine the tensor structures.  We next turn our attention to these objects.


\subsection{Tensor Structures and Three-Point Correlation Functions}

The tensor structures are the remaining essential ingredients required to construct three-point correlation functions.  Since these are purely group theoretic objects, only the irreducible representations of the quasi-primary operators are needed to determine them.  As mentioned in Section \ref{SecTensor}, the tensor structures can always be orthonormalized starting from an arbitrary basis of such structures, which is usually quite simple to generate.  In the following, the irreducible representations of the three quasi-primary operators appearing in the three-point functions will be delimited by square brackets as in $[\boldsymbol{N}_i,\boldsymbol{N}_j,\boldsymbol{N}_k]$, and the quantity of interest appearing in the correlation functions \eqref{EqCFSub} will be
\eqn{\lambda_{\boldsymbol{N}_k}\bar{J}_{12;3}^{(d,h_{ijk},n_a,\Delta_k,\boldsymbol{N}_k)}\cdot\tCF{a}{i}{j}{k}{1}{2}.}[EqThreePt]
Moreover, the contractions between the $\bar{J}$-function and the tensor structures will be simplified using the contiguous relations \eqref{EqCont}.


\subsubsection{Scalar-Scalar-\texorpdfstring{$($}{(}Symmetric-Traceless\texorpdfstring{$)$}{)}}

For the $[\boldsymbol{0},\boldsymbol{0},\ell\boldsymbol{e}_1]$ three-point correlation functions, the only possible tensor structure must contract the free indices on $\ell\boldsymbol{e}_1$ with the differential operator.  Since $\A_{12C_1}^{\phantom{12C_1}D_1}\cdots\A_{12C_\ell}^{\phantom{12C_\ell}D_\ell}$ explicitly contracts all indices without vanishing, the basis tensor structure can be obtained by applying \eqref{EqPTS} on this quantity.  Doing so results in the tensor structure $\tCF{}{i}{j}{k}{1}{2}=\lambda_{\ell\boldsymbol{e}_1}\hat{\mathcal{P}}_{12}^{\ell\boldsymbol{e}_1}(g)^\ell$, where $\lambda_{\ell\boldsymbol{e}_1}=\sqrt{\ell!/[(d+2\ell-2)(d-1)_{\ell-1}]}$ is the normalization constant appearing in \eqref{EqOrthonorm}.  It is important to observe here that any other nonvanishing choice of contractions leads to the same tensor structure.  Indeed, contractions like $\A_{12C_1}^{\phantom{12C_1}D_2}\A_{12C_2}^{\phantom{12C_2}D_1}\A_{12C_3}^{\phantom{12C_3}D_3}\cdots\A_{12C_\ell}^{\phantom{12C_\ell}D_\ell}$ are equivalent, while contractions such as $\A_{12C_1C_2}\A_{12}^{D_1D_2}\A_{12C_3}^{\phantom{12C_3}D_3}\cdots\A_{12C_\ell}^{\phantom{12C_\ell}D_\ell}$ vanish due to the tracelessness of the irreducible representations of the Lorentz group.

We therefore obtain
\eqn{\lambda_{\ell\boldsymbol{e}_1}(\bar{J}_{12;3}^{(d,h_{ijk},\ell,\Delta_k,\ell\boldsymbol{e}_1)}\cdot\tCF{}{i}{j}{k}{1}{2})_{\{C\}}=(\lambda_{\ell\boldsymbol{e}_1})^2(\hat{\mathcal{P}}_{12}^{\ell\boldsymbol{e}_1})_{\{C'\}}^{\phantom{\{C'\}}\{D'\}}(g_{D'D})^\ell(\bar{J}_{12;3}^{(d,h_{ijk},\ell,\Delta_k,\ell\boldsymbol{e}_1)})_{\{C\}}^{\phantom{\{C\}}\{C'\}\{D\}},}
where the hatted projection operator and the $\bar{J}$-function are \eqref{EqPle1} and \eqref{EqJble1} respectively.  Focusing on $\boldsymbol{0}$ and $\boldsymbol{e}_1$ exchange \eqref{EqJb0ande1}, the three-point correlation function contributions \eqref{EqThreePt} are
\eqna{
\lambda_{\boldsymbol{0}}(\bar{J}_{12;3}^{(d,h_{ijk},0,\Delta_k,\boldsymbol{0})}\cdot\tCF{}{i}{j}{k}{1}{2})&=\bar{I}_{12}^{(d,h_{ijk},0,\Delta_k)}=(-2)^{h_{ijk}}(\Delta_k)_{h_{ijk}}(\Delta_k+1-d/2)_{h_{ijk}},\\
\lambda_{\boldsymbol{e}_1}(\bar{J}_{12;3}^{(d,h_{ijk},1,\Delta_k,\boldsymbol{e}_1)}\cdot\tCF{}{i}{j}{k}{1}{2})_C&=\frac{1}{d}\A_{12C'D}\left(g_C^{\phantom{C}C'}\bar{I}_{12}^{(d,h_{ijk}-1/2,1;\Delta_k)D}-\bar{\eta}_{1C}\bar{I}_{12}^{(d,h_{ijk}-3/2,2;\Delta_k)C'D}\right.\\
&\phantom{=}\qquad\left.-\bar{\eta}_3^{C'}\bar{I}_{12}^{(d,h_{ijk}-3/2,2;\Delta_k+1)}{}_C^{\phantom{C}D}+\bar{I}_{12}^{(d,h_{ijk}-5/2,3;\Delta_k+1)}{}_C^{\phantom{C}C'D}\right)\\
&=\frac{1}{d}\left[\bar{I}_{12}^{(d,h_{ijk}-1/2,1;\Delta_k+1)}{}_C-\bar{\eta}_{2C}\bar{I}_{12}^{(d,h_{ijk}+1/2,0;\Delta_k)}\right.\\
&\phantom{=}\qquad\left.+(h_{ijk}+1/2)(2h_{ijk}-1+d)\bar{I}_{12}^{(d,h_{ijk}-3/2,1;\Delta_k+1)}{}_C\right]\\
&=\frac{2}{d}(-2)^{h_{ijk}-1/2}(h_{ijk}+1/2)(d-1-\Delta_k)\\
&\phantom{=}\qquad\times(\Delta_k+1)_{h_{ijk}-1/2}(\Delta_k+1-d/2)_{h_{ijk}-1/2}\bar{\eta}_{2C}.
}
Here we have discarded all contributions containing $\bar{\eta}_{1C}$ and $\bar{\eta}_{3C}$, since they are contracted with the half-projector for the exchanged quasi-primary operator, which is transverse with respect to these coordinates, as evident from \eqref{EqCFSub}.  As a consequence, such contributions vanish and can be disregarded altogether.  In addition, the contiguous relations \eqref{EqCont} and the definition \eqref{EqIb} were used to further simplify the final results.

Following the same procedure, we find that the contribution \eqref{EqThreePt} to the three-point correlation function $[\boldsymbol{0},\boldsymbol{0},\ell\boldsymbol{e}_1]$ is given by
\eqna{
\lambda_{\ell\boldsymbol{e}_1}(\bar{J}_{12;3}^{(d,h_{ijk},\ell,\Delta_k,\ell\boldsymbol{e}_1)}\cdot\tCF{}{i}{j}{k}{1}{2})_{C_\ell\cdots C_1}&=\frac{(-2)^{h_{ijk}-\ell/2}2^\ell\ell!(h_{ijk}-\ell/2+1)_\ell(d-1-\Delta_k)_\ell}{(d+2\ell-2)(d-1)_{\ell-1}}\\
&\phantom{=}\qquad\times(\Delta_k+\ell)_{h_{ijk}-\ell/2}(\Delta_k+1-d/2)_{h_{ijk}-\ell/2}\bar{\eta}_{2C_\ell}\cdots\bar{\eta}_{2C_1}.
}[Eq00le1]

To further verify the formalism introduced in \cite{Fortin:2019fvx,Fortin:2019dnq}, we project the three-point correlation functions to position space.  We have
\eqna{
\Vev{\Op{i}{1}\Op{j}{2}\Op{k}{3}}&=\frac{(-2)^{h_{ijk}-\ell/2}2^\ell\ell!(h_{ijk}-\ell/2+1)_\ell(d-1-\Delta_k)_\ell}{(d+2\ell-2)(d-1)_{\ell-1}}(\Delta_k+\ell)_{h_{ijk}-\ell/2}\\
&\phantom{=}\qquad\times\frac{(\Delta_k+1-d/2)_{h_{ijk}-\ell/2}\cCF{}{i}{j}{k}(\mathcal{T}_{31}^{\ell\boldsymbol{e}_1}\Gamma)^{C_1\cdots C_\ell}\bar{\eta}_{2C_\ell}\cdots\bar{\eta}_{2C_1}}{\ee{1}{2}{\frac{1}{2}(\Delta_i+\Delta_j-\Delta_k)}\ee{1}{3}{\frac{1}{2}(\Delta_i-\Delta_j+\Delta_k-\ell)}\ee{2}{3}{\frac{1}{2}(-\Delta_i+\Delta_j+\Delta_k)}},
}[EqThreePt00le1]
using \eqref{EqCFSub} and \eqref{Eq00le1}.  Since the position space quasi-primary operators are obtained by projecting to the first half of the embedding space spinor indices and multiplying by the proper homogeneity factor, $\mathcal{O}^{(x)}(x)=(-\eta^{d+1}+\eta^{d+2})^{\tau}\Op{+}{}$, the three-point correlation functions for $\ell=0,1$ are therefore proportional to
\begingroup\makeatletter\def\f@size{10}\check@mathfonts\def\maketag@@@#1{\hbox{\m@th\large\normalfont#1}}%
\eqna{
\Vev{\mathcal{O}_i^{(x)}(x_1)\mathcal{O}_j^{(x)}(x_2)\mathcal{O}_k^{(x)}(x_3)}&\propto\cCF{}{i}{j}{k}\left[-(x_1-x_2)^2/2\right]^{-\frac{1}{2}(\Delta_i+\Delta_j-\Delta_k)}\left[-(x_1-x_3)^2/2\right]^{-\frac{1}{2}(\Delta_i-\Delta_j+\Delta_k)}\\
&\phantom{\propto}\qquad\times\left[-(x_2-x_3)^2/2\right]^{-\frac{1}{2}(-\Delta_i+\Delta_j+\Delta_k)},\\
\Vev{\mathcal{O}_i^{(x)}(x_1)\mathcal{O}_j^{(x)}(x_2)\mathcal{O}_{k\gamma_1\gamma_2}^{(x)}(x_3)}&\propto\cCF{}{i}{j}{k}(\gamma^\mu C^{-1})_{\gamma_1\gamma_2}R_\mu(x_1,x_2|x_3)\left[-(x_1-x_2)^2/2\right]^{-\frac{1}{2}(\Delta_i+\Delta_j-\Delta_k)}\\
&\phantom{\propto}\qquad\times\left[-(x_1-x_3)^2/2\right]^{-\frac{1}{2}(\Delta_i-\Delta_j+\Delta_k)}\left[-(x_2-x_3)^2/2\right]^{-\frac{1}{2}(-\Delta_i+\Delta_j+\Delta_k)},
}
\endgroup
which indeed match the expected results in position space.  In the last equation, the projection led to
\eqna{
R_\mu(x_1,x_2|x_3)&=\frac{[-(\A_{31}\cdot\bar{\eta}_2)^{d+1}+(\A_{31}\cdot\bar{\eta}_2)^{d+2}]\eta_{3\mu}-(-\eta_3^{d+1}+\eta_3^{d+2})(\A_{31}\cdot\bar{\eta}_2)_\mu}{\sqrt{-2}(-\eta_3^{d+1}+\eta_3^{d+2})}\\
&=\sqrt{-1/2}[-\bar{\eta}_2^{d+1}+\bar{\eta}_2^{d+2}+\bar{\eta}_3^{d+1}-\bar{\eta}_3^{d+2}+\bar{\eta}_1^{d+1}-\bar{\eta}_1^{d+2}]x_{3\mu}-\sqrt{-1/2}(\bar{\eta}_{2\mu}-\bar{\eta}_{3\mu}-\bar{\eta}_{1\mu})\\
&=\frac{|x_2-x_3|(x_{1\mu}-x_{3\mu})}{|x_1-x_2||x_1-x_3|}-\frac{|x_1-x_3|(x_{2\mu}-x_{3\mu})}{|x_1-x_2||x_2-x_3|},
}
contracted with the gamma matrix.  Generalizing to $\ell\boldsymbol{e}_1$ shows that the three-point correlation functions \eqref{EqThreePt00le1} project to the well-known results found in the literature (see \textit{e.g.} \cite{Simmons-Duffin:2016gjk}), validating the formalism introduced in \cite{Fortin:2019fvx,Fortin:2019dnq} for three-point correlation functions.


\subsubsection{Scalar-Vector-Vector and Permutations}

We now consider the scalar-vector-vector three-point function.  In the case of $[\boldsymbol{0},\boldsymbol{e}_1,\boldsymbol{e}_1]$ ($[\boldsymbol{0},2\boldsymbol{e}_1,2\boldsymbol{e}_1]$ in three spacetime dimensions), the number of tensor structures is two when $d>3$ (three when $d=3$), as exemplified by the tensor product decompositions
\eqn{
\begin{gathered}
d>3:\qquad\boldsymbol{0}\otimes\boldsymbol{e}_1\otimes\boldsymbol{e}_1=2\boldsymbol{e}_1\oplus\boldsymbol{e}_2\oplus\boldsymbol{0}, \\
d=3:\qquad\boldsymbol{0}\otimes2\boldsymbol{e}_1\otimes2\boldsymbol{e}_1=4\boldsymbol{e}_1\oplus2\boldsymbol{e}_1\oplus\boldsymbol{0}.
\end{gathered}
}
Clearly, the tensor structures corresponding to $2\boldsymbol{e}_1$ ($4\boldsymbol{e}_1$ in three spacetime dimensions) and $\boldsymbol{0}$ are made out of $\A_{12BD_1}\A_{12CD_2}$ and $\A_{12BC}$, respectively.  In three spacetime dimensions, the remaining tensor structure, corresponding to $2\boldsymbol{e}_1$, comes from the epsilon tensor and is constructed from $\epsilon_{12BCD}$.  Therefore, the tensor structures satisfying \eqref{EqPTS} are
\eqn{
\begin{gathered}
d>3:\qquad(\tCF{1}{i}{j}{k}{1}{2})_{BC'D_2D_1}=\sqrt{\frac{2}{(d-1)(d+2)}}\left[\A_{12B(D_1}\A_{12C'D_2)}-\frac{1}{d}\A_{12BC'}\A_{12D_1D_2}\right],\\
(\tCF{2}{i}{j}{k}{1}{2})_{BC'}=\frac{1}{\sqrt{d}}\A_{12BC'},\\
d=3:\qquad(\tCF{1}{i}{j}{k}{1}{2})_{BC'D_2D_1}=\frac{1}{\sqrt{5}}\left[\A_{12B(D_1}\A_{12C'D_2)}-\frac{1}{3}\A_{12BC'}\A_{12D_1D_2}\right],\\
(\tCF{2}{i}{j}{k}{1}{2})_{BC'}=\frac{1}{\sqrt{3}}\A_{12BC'},\qquad(\tCF{3}{i}{j}{k}{1}{2})_{BC'D}=\frac{1}{\sqrt{6}}\epsilon_{12BC'D},
\end{gathered}
}[EqTS0e1e1]
which are automatically orthogonal due to the grading and were normalized following \eqref{EqOrthonorm}.

Using the tensor structures \eqref{EqTS0e1e1} and the $\bar{J}$-function \eqref{EqJb0ande1}, we may readily determine the contributions \eqref{EqThreePt} to the three-point correlation function $[\boldsymbol{0},\boldsymbol{e}_1,\boldsymbol{e}_1]$ ($[\boldsymbol{0},2\boldsymbol{e}_1,2\boldsymbol{e}_1]$ in $d=3$) to be
\eqna{
\lambda_{\boldsymbol{e}_1}(\bar{J}_{12;3}^{(d,h_{ijk},2,\Delta_k,\boldsymbol{e}_1)}\cdot\tCF{1}{i}{j}{k}{1}{2})_{BC}&=\frac{8(-2)^{h_{ijk}-1}(\Delta_k+1)_{h_{ijk}-1}(\Delta_k+1-d/2)_{h_{ijk}-1}(h_{ijk})_2}{d\sqrt{d(d-1)(d+2)/2}}\\
&\phantom{=}\qquad\times\{[\Delta_k(\Delta_k+h_{ijk}-d)-h_{ijk}-d/2+d^2/2]g_{BC}\\
&\phantom{=}\qquad+(d/2-1)(\Delta_k+h_{ijk})(\Delta_k-d)\bar{\eta}_{3B}\bar{\eta}_{2C}\},\\
\lambda_{\boldsymbol{e}_1}(\bar{J}_{12;3}^{(d,h_{ijk},0,\Delta_k,\boldsymbol{e}_1)}\cdot\tCF{2}{i}{j}{k}{1}{2})_{BC}&=\frac{1}{d}(-2)^{h_{ijk}}(\Delta_k+1)_{h_{ijk}-1}(\Delta_k+1-d/2)_{h_{ijk}-1}\\
&\phantom{=}\qquad\times\{[\Delta_k(\Delta_k+h_{ijk}-d/2)-h_{ijk}]g_{BC}\\
&\phantom{=}\qquad-(\Delta_k+h_{ijk})(\Delta_k-d/2)\bar{\eta}_{3B}\bar{\eta}_{2C}\},
}[Eq0e1e14d]
in spacetime dimensions larger than three and
\eqna{
\lambda_{2\boldsymbol{e}_1}(\bar{J}_{12;3}^{(d,h_{ijk},2,\Delta_k,2\boldsymbol{e}_1)}\cdot\tCF{1}{i}{j}{k}{1}{2})_{BC}&=\frac{8(-2)^{h_{ijk}-1}(\Delta_k+1)_{h_{ijk}-1}(\Delta_k-1/2)_{h_{ijk}-1}(h_{ijk})_2}{3\sqrt{15}}\\
&\phantom{=}\qquad\times\{[\Delta_k(\Delta_k+h_{ijk}-3)-h_{ijk}+3]g_{BC}\\
&\phantom{=}\qquad+(1/2)(\Delta_k+h_{ijk})(\Delta_k-3)\bar{\eta}_{3B}\bar{\eta}_{2C}\},\\
\lambda_{2\boldsymbol{e}_1}(\bar{J}_{12;3}^{(d,h_{ijk},0,\Delta_k,2\boldsymbol{e}_1)}\cdot\tCF{2}{i}{j}{k}{1}{2})_{BC}&=\frac{1}{3}(-2)^{h_{ijk}}(\Delta_k+1)_{h_{ijk}-1}(\Delta_k-1/2)_{h_{ijk}-1}\\
&\phantom{=}\qquad\times\{[\Delta_k(\Delta_k+h_{ijk}-3/2)-h_{ijk}]g_{BC}\\
&\phantom{=}\qquad-(\Delta_k+h_{ijk})(\Delta_k-3/2)\bar{\eta}_{3B}\bar{\eta}_{2C}\},\\
\lambda_{2\boldsymbol{e}_1}(\bar{J}_{12;3}^{(d,h_{ijk},1,\Delta_k,2\boldsymbol{e}_1)}\cdot\tCF{3}{i}{j}{k}{1}{2})_{BC}&=-\frac{\sqrt{2}}{3}(-2)^{h_{ijk}-1/2}(\Delta_k+1)_{h_{ijk}-1/2}(\Delta_k+1-d/2)_{h_{ijk}-1/2}\\
&\phantom{=}\qquad\times(\Delta_k-1)(h_{ijk}+1/2)\epsilon_{12BCD}\bar{\eta}_3^D,
}[Eq0e1e13d]
in three spacetime dimensions.  All contributions that obviously vanish when contracted with the $\mathcal{T}_{ij}^{\boldsymbol{N}}\Gamma$ were discarded.

From \eqref{Eq0e1e14d} and \eqref{Eq0e1e13d}, it is straightforward to see that all contributions are linearly independent, as expected.  However, these are not the simplest linear combinations at the level of the three-point correlation functions, although they were quite natural from the perspective of the OPE.  This can be understood from the appearance of the third embedding space coordinate in three-point correlation functions, as emphasized in \cite{Fortin:2019dnq}.

To somewhat elucidate the permutation properties \eqref{EqPerm}, it is of interest to study the $[\boldsymbol{e}_1,\boldsymbol{e}_1,\boldsymbol{0}]$ correlation function ($[2\boldsymbol{e}_1,2\boldsymbol{e}_1,\boldsymbol{0}]$ in $d=3$), where the first and last quasi-primary operators have been interchanged.  From \eqref{EqTS0e1e1}, the tensor structures are
\eqn{
\begin{gathered}
d>3:\qquad(\tCF{1}{k}{j}{i}{1}{2})_{ABD_2D_1}=\sqrt{\frac{2}{(d-1)(d+2)}}\left[\A_{12A(D_1}\A_{12BD_2)}-\frac{1}{d}\A_{12AB}\A_{12D_1D_2}\right],\\
(\tCF{2}{k}{j}{i}{1}{2})_{AB}=\frac{1}{\sqrt{d}}\A_{12AB},\\
d=3:\qquad(\tCF{1}{k}{j}{i}{1}{2})_{ABD_2D_1}=\frac{1}{\sqrt{5}}\left[\A_{12A(D_1}\A_{12BD_2)}-\frac{1}{3}\A_{12AB}\A_{12D_1D_2}\right],\\
(\tCF{2}{k}{j}{i}{1}{2})_{AB}=\frac{1}{\sqrt{3}}\A_{12AB},\qquad(\tCF{3}{k}{j}{i}{1}{2})_{ABD}=\frac{1}{\sqrt{6}}\epsilon_{12ABD},
\end{gathered}
}[EqTSe1e10]
and the contributions to the three-point correlation functions can be expressed from \eqref{EqTSe1e10} and \eqref{EqJb0ande1} as
\eqna{
\lambda_{\boldsymbol{0}}(\bar{J}_{12;3}^{(d,h_{kji},2,\Delta_i,\boldsymbol{0})}\cdot\tCF{1}{k}{j}{i}{1}{2})_{AB}&=\frac{2(-2)^{h_{kji}+1}(\Delta_i)_{h_{kji}+1}(\Delta_i+1-d/2)_{h_{kji}-1}(h_{kji})_2}{d\sqrt{(d-1)(d+2)/2}}\\
&\phantom{=}\qquad\times[g_{AB}+(d/2)\bar{\eta}_{3A}\bar{\eta}_{3B}],\\
\lambda_{\boldsymbol{0}}(\bar{J}_{12;3}^{(d,h_{kji},0,\Delta_i,\boldsymbol{0})}\cdot\tCF{2}{k}{j}{i}{1}{2})_{AB}&=\frac{1}{\sqrt{d}}(-2)^{h_{kji}}(\Delta_i)_{h_{kji}}(\Delta_i+1-d/2)_{h_{kji}}g_{AB},
}[Eqe1e104d]
in spacetime dimensions larger than three and
\eqna{
\lambda_{\boldsymbol{0}}(\bar{J}_{12;3}^{(d,h_{kji},2,\Delta_i,\boldsymbol{0})}\cdot\tCF{1}{k}{j}{i}{1}{2})_{AB}&=\frac{2(-2)^{h_{kji}+1}(\Delta_i)_{h_{kji}+1}(\Delta_i-1/2)_{h_{kji}-1}(h_{kji})_2}{3\sqrt{5}}\\
&\phantom{=}\qquad\times[g_{AB}+(3/2)\bar{\eta}_{3A}\bar{\eta}_{3B}],\\
\lambda_{\boldsymbol{0}}(\bar{J}_{12;3}^{(d,h_{kji},0,\Delta_i,\boldsymbol{0})}\cdot\tCF{2}{k}{j}{i}{1}{2})_{AB}&=\frac{1}{\sqrt{3}}(-2)^{h_{kji}}(\Delta_i)_{h_{kji}}(\Delta_i-1/2)_{h_{kji}}g_{AB},\\
\lambda_{\boldsymbol{0}}(\bar{J}_{12;3}^{(d,h_{kji},1,\Delta_i,\boldsymbol{0})}\cdot\tCF{3}{k}{j}{i}{1}{2})_{AB}&=-\sqrt{\frac{2}{3}}(-2)^{h_{kji}-1/2}(\Delta_i)_{h_{kji}-1/2}(\Delta_i+1-d/2)_{h_{kji}-1/2}\\
&\phantom{=}\qquad\times(\Delta_i+h_{kji}-1/2)(h_{kji}+1/2)\epsilon_{12ABD}\bar{\eta}_3^D,
}[Eqe1e103d]
in three spacetime dimensions.

Focusing on spacetime dimensions larger than three for simplicity, \eqref{Eqe1e104d} implies that
\eqna{
&\cCF{1}{k}{j}{i}\frac{2(-2)^{h_{kji}+1}(\Delta_i)_{h_{kji}+1}(\Delta_i+1-d/2)_{h_{kji}-1}(h_{kji})_2}{d\sqrt{(d-1)(d+2)/2}}[g_{CB}+(d/2-1)\bar{\eta}_{2C}\bar{\eta}_{3B}]\\
&\phantom{=}\qquad+\cCF{2}{k}{j}{i}\frac{1}{\sqrt{d}}(-2)^{h_{kji}}(\Delta_i)_{h_{kji}}(\Delta_i+1-d/2)_{h_{kji}}(g_{CB}-\bar{\eta}_{2C}\bar{\eta}_{3B})\\
&=\cCF{1}{i}{j}{k}\frac{8(-2)^{h_{ijk}-1}(\Delta_k+1)_{h_{ijk}-1}(\Delta_k+1-d/2)_{h_{ijk}-1}(h_{ijk})_2}{d\sqrt{d(d-1)(d+2)/2}}\\
&\phantom{=}\qquad\times\{[\Delta_k(\Delta_k+h_{ijk}-d)-h_{ijk}-d/2+d^2/2]g_{BC}\\
&\phantom{=}\qquad+(d/2-1)(\Delta_k+h_{ijk})(\Delta_k-d)\bar{\eta}_{3B}\bar{\eta}_{2C}\}\\
&\phantom{=}\qquad+\cCF{2}{i}{j}{k}\frac{1}{d}(-2)^{h_{ijk}}(\Delta_k+1)_{h_{ijk}-1}(\Delta_k+1-d/2)_{h_{ijk}-1}\\
&\phantom{=}\qquad\times\{[\Delta_k(\Delta_k+h_{ijk}-d/2)-h_{ijk}]g_{BC}\\
&\phantom{=}\qquad-(\Delta_k+h_{ijk})(\Delta_k-d/2)\bar{\eta}_{3B}\bar{\eta}_{2C}\}
}
after exchanging $\eta_1\leftrightarrow\eta_3$, extracting the proper $\A$-metrics, and relabelling the dummy indices, as expected from \eqref{EqPerm}.  Hence, we find that the three-point function OPE coefficients are related to one another under index permutations, albeit in quite an intricate way when the three-point correlation functions are computed from the OPE.


\subsubsection{Vector-Vector-Vector in Three Spacetime Dimensions}

The previous example should make it clear that once the tensor structures are obtained, it is straightforward to compute the associated three-point correlation functions.  To illustrate the algorithm for the construction of tensor structures discussed in Section \ref{SecTensor}, we present here a more contrived example with several tensor structures originating from the same symmetric-traceless irreducible representation, namely $[2\boldsymbol{e}_1,2\boldsymbol{e}_1,2\boldsymbol{e}_1]$ in three spacetime dimensions.  We restrict attention to the tensor structures alone and perform the analysis in three spacetime dimensions, in order to include several parity-violating contributions.

The tensor product decomposition of interest is
\eqn{2\boldsymbol{e}_1\otimes2\boldsymbol{e}_1\otimes2\boldsymbol{e}_1=6\boldsymbol{e}_1\oplus\boldsymbol{0}\oplus2\boldsymbol{e}_1\oplus2\boldsymbol{e}_1\oplus2\boldsymbol{e}_1\oplus4\boldsymbol{e}_1\oplus4\boldsymbol{e}_1.}
This contains seven symmetric-traceless irreducible representations, and therefore seven tensor structures, which we will enumerate from left to right.  Constructing the lone tensor structures $6\boldsymbol{e}_1$ and $\boldsymbol{0}$ is simple.  They correspond to
\eqn{(\tCF{1}{i}{j}{k}{1}{2})_{ABC'D_3D_2D_1}=\frac{1}{\sqrt{7}}(\hat{\mathcal{P}}^{6\boldsymbol{e}_1})_{ABC'}^{\phantom{ABC}D'_1D'_2D'_3}g_{D'_3D_3}g_{D'_2D_2}g_{D'_1D_1},\qquad(\tCF{2}{i}{j}{k}{1}{2})_{ABC'}=\frac{1}{\sqrt{6}}\epsilon_{12ABC'}.}
A direct computation shows that these satisfy \eqref{EqPTS}, and their normalization constants can be easily obtained using \eqref{EqOrthonorm}.  They are obviously orthogonal among themselves and all the other tensor structures due to the grading.

For the three tensor structures corresponding to the three $2\boldsymbol{e}_1$, a basis can be obtained by writing down all possible contractions, giving
\eqn{(\vCF{3}{i}{j}{k}{1}{2})_{ABC'D}=\A_{12AB}\A_{12C'D},\quad(\vCF{4}{i}{j}{k}{1}{2})_{ABC'D}=\A_{12AC'}\A_{12BD},\quad(\vCF{5}{i}{j}{k}{1}{2})_{ABC'D}=\A_{12BC'}\A_{12AD}.}
The tensor structures automatically satisfy \eqref{EqPTS} simply because the projection operators are trivial for $2\boldsymbol{e}_1$.  They can be rotated to orthogonalize the basis using any known technique, for example giving
\eqn{
\begin{gathered}
(\tCF{3}{i}{j}{k}{1}{2})_{ABC'D}=\frac{1}{3\sqrt{5}}(\A_{12AB}\A_{12C'D}+\A_{12AC'}\A_{12BD}+\A_{12BC'}\A_{12AD}),\\
(\tCF{4}{i}{j}{k}{1}{2})_{ABC'D}=\frac{1}{2\sqrt{3}}(\A_{12AB}\A_{12C'D}-\A_{12AC'}\A_{12BD}),\\
(\tCF{5}{i}{j}{k}{1}{2})_{ABC'D}=\frac{1}{6}(\A_{12AB}\A_{12C'D}+\A_{12AC'}\A_{12BD}-2\A_{12BC'}\A_{12AD}).
\end{gathered}
}
Again, the normalization constants are trivially computed from \eqref{EqOrthonorm}.

Finally, the remaining two tensor structures for the two $4\boldsymbol{e}_1$ are slightly more difficult to generate.  They are parity odd, like $\tCF{2}{i}{j}{k}{1}{2}$, due to the presence of the epsilon tensor.  With five vector indices, there are three such potential combinations built from $\epsilon_{12ABD_1}\A_{12C'D_2}$, $\epsilon_{12AC'D_1}\A_{12BD_2}$, and $\epsilon_{12BC'D_1}\A_{12AD_2}$; however, there must be only two linearly-independent combinations.  Implementing the condition \eqref{EqPTS} gives
\eqn{
\begin{gathered}
(\vCF{6}{i}{j}{k}{1}{2})_{ABC'D_2D_1}=\epsilon_{12AB(D_1}\A_{12D_2)C'}-\frac{1}{3}\epsilon_{12ABC'}\A_{12D_1D_2},\\
(\vCF{7}{i}{j}{k}{1}{2})_{ABC'D_2D_1}=\epsilon_{12AC'(D_1}\A_{12D_2)B}+\frac{1}{3}\epsilon_{12ABC'}\A_{12D_1D_2},\\
(\vCF{8}{i}{j}{k}{1}{2})_{ABC'D_2D_1}=\epsilon_{12BC'(D_1}\A_{12D_2)A}-\frac{1}{3}\epsilon_{12ABC'}\A_{12D_1D_2}.
\end{gathered}
}
Here it is not clear that one of the linear tensor structures is linearly dependent, although the condition \eqref{EqPTS} has been enforced.  This situation is reminiscent of the construction of the same tensor structures in position space.  To see this, it is helpful to compute the matrix $M$ of inner products for these three tensor structures as suggested in Section \ref{SecTensor}.  Using $\epsilon_{12ABC}\epsilon_{21}^{ABC}=6$, we get
\eqn{M=5\left(\begin{array}{ccc}2&1&-1\\1&2&1\\-1&1&2\end{array}\right),}
which has rank two, implying only two independent tensor structures.  This matrix of products is exactly the same as that computed from the corresponding tensor structures in position space.  One possible set of orthonormalized structures is thus given by
\eqn{
\begin{gathered}
(\tCF{6}{i}{j}{k}{1}{2})_{ABC'D_2D_1}=\frac{1}{\sqrt{30}}[\epsilon_{12AB(D_1}\A_{12D_2)C'}+\epsilon_{12AC'(D_1}\A_{12D_2)B}],\\
(\tCF{7}{i}{j}{k}{1}{2})_{ABC'D_2D_1}=\frac{1}{\sqrt{90}}[\epsilon_{12AB(D_1}\A_{12D_2)C'}-\epsilon_{12AC'(D_1}\A_{12D_2)B}-2\epsilon_{12BC'(D_1}\A_{12D_2)A}],
\end{gathered}
}
where the last eigenvector, the extra tensor structure
\eqn{(\tCF{8}{i}{j}{k}{1}{2})_{ABC'D_2D_1}=\epsilon_{12AB(D_1}\A_{12D_2)C'}-\epsilon_{12AC'(D_1}\A_{12D_2)B}+\epsilon_{12BC'(D_1}\A_{12D_2)A}-\epsilon_{12ABC'}\A_{12D_1D_2},}[EqNTS]
has zero norm and should therefore be discarded.  In fact, \eqref{EqNTS} should not merely have a vanishing norm; it should actually vanish identically.  In position space, it can be shown with the help of some identities that the analog of \eqref{EqNTS} is actually identically zero.  In embedding space, where there are more choices for the dummy indices, such a conclusion is not apparent at first.  However, a direct computation reveals that \eqref{EqNTS} vanishes on the light cone, in complete agreement with the position space result.  Hence, on the light cone, the extra tensor structure with zero norm \eqref{EqNTS} vanishes and can therefore be forgotten altogether, as expected from position space.

Finally, we want to stress that the analysis of embedding space tensor structures is completely analogous to the analysis of position space structures, as exemplified by the discussion above, demonstrating the convenience of the formalism introduced in \cite{Fortin:2019fvx,Fortin:2019dnq}.  At this point, it is a trivial matter to compute the three-point correlation function contributions \eqref{EqThreePt} from the tensor structures obtained above to determine the full three-point correlation function \eqref{EqCFSub}.


\subsubsection{Scalar-Spinor-Spinor and Permutations}

To illustrate that the formalism applies uniformly to any irreducible representation, we now investigate three-point correlation functions for quasi-primary operators in spinor representations.  For simplicity, we focus on odd spacetime dimensions, although the even-dimensional case is similar.

We consider the $[\boldsymbol{0},\boldsymbol{e}_r,\boldsymbol{e}_r]$ three-point correlation function, for which the tensor product decomposition gives
\eqn{\boldsymbol{0}\otimes\boldsymbol{e}_r\otimes\boldsymbol{e}_r\supset\boldsymbol{e}_1\oplus\boldsymbol{0}.}
The omitted irreducible representations on the right-hand side above are not symmetric-traceless irreducible representations and therefore do not correspond to any three-point functions (in three spacetime dimensions, the vector irreducible representation on the right-hand side of the tensor product decomposition would be $2\boldsymbol{e}_1$ instead of $\boldsymbol{e}_1$).  Thus, there are two tensor structures, given by
\eqn{(\tCF{1}{i}{j}{k}{1}{2})_{bc'D}=\frac{1}{\sqrt{2^{r+1}d}}(\Gamma_{12D}C_\Gamma^{-1})_{bc'},\qquad(\tCF{2}{i}{j}{k}{1}{2})_{bc'}=\frac{1}{\sqrt{2^{r+1}}}(C_\Gamma^{-1})_{bc'},}
for the $\boldsymbol{e}_1$ and the $\boldsymbol{0}$ irreducible representations, respectively.  Using the $\bar{J}$-function \eqref{EqJber} with $\lambda_{\boldsymbol{e}_r}=1/\sqrt{2^{r+1}}$ leads to
\eqna{
\lambda_{\boldsymbol{e}_r}(\bar{J}_{12;3}^{(d,h_{ijk},1,\Delta_k,\boldsymbol{e}_r)}\cdot\tCF{1}{i}{j}{k}{1}{2})_{bc}&=\frac{(-1)^{r(r+1)/2}}{2^{r+1}\sqrt{d}}(\bar{\eta}_3\cdot\Gamma\Gamma_{D_0}\Gamma_{12D_1}C_\Gamma^{-1})_{cb}\bar{I}_{12}^{(d,h_{ijk}-3/2,2;\Delta_k+1/2)D_0D_1}\\
&=\frac{(-1)^{r(r+1)/2}}{2^{r+1}\sqrt{d}}2(-2)^{h_{ijk}+1/2}(\Delta_k+1/2)_{h_{ijk}-1/2}\\
&\phantom{=}\qquad\times(\Delta_k+1/2-d/2)_{h_{ijk}+1/2}(h_{ijk}+1/2)(\bar{\eta}_3\cdot\Gamma C_\Gamma^{-1})_{cb},
}[Eq0erer1]
and
\eqna{
\lambda_{\boldsymbol{e}_r}(\bar{J}_{12;3}^{(d,h_{ijk},0,\Delta_k,\boldsymbol{e}_r)}\cdot\tCF{2}{i}{j}{k}{1}{2})_{bc}&=\frac{(-1)^{(r+1)(r+2)/2}}{2^{r+1}}(\bar{\eta}_3\cdot\Gamma\Gamma_DC_\Gamma^{-1})_{cb}\bar{I}_{12}^{(d,h_{ijk}-1,1;\Delta_k+1/2)D}\\
&=\frac{(-1)^{(r+1)(r+2)/2}}{2^{r+1}}(-2)^{h_{ijk}}(\Delta_k+1/2)_{h_{ijk}}(\Delta_k+1/2-d/2)_{h_{ijk}}\\
&\phantom{=}\qquad\times(\bar{\eta}_3\cdot\Gamma\bar{\eta}_2\cdot\Gamma C_\Gamma^{-1})_{cb},
}[Eq0erer2]
where all contributions that vanish upon contraction with the half-projectors appearing in \eqref{EqCFSub} were discarded, and trivial algebraic simplifications were made.  The two three-point correlation function contributions \eqref{Eq0erer1} and \eqref{Eq0erer2} lead to the expected results, as can be seen by using the explicit form of the half-projectors for spinor representations $\mathcal{T}_{ij}^{\boldsymbol{e}_r}\Gamma=\bar{\eta}_i\cdot\Gamma\bar{\eta}_j\cdot\Gamma/\sqrt{2}$.

Turning to the permuted three-point correlation function $[\boldsymbol{e}_r,\boldsymbol{e}_r,\boldsymbol{0}]$, the tensor structures are
\eqn{(\tCF{1}{k}{j}{i}{1}{2})_{abD}=\frac{1}{\sqrt{2^{r+1}d}}(\Gamma_{12D}C_\Gamma^{-1})_{ab},\qquad(\tCF{2}{k}{j}{i}{1}{2})_{ab}=\frac{1}{\sqrt{2^{r+1}}}(C_\Gamma^{-1})_{ab},}
which instead result in
\eqna{
\lambda_{\boldsymbol{0}}(\bar{J}_{12;3}^{(d,h_{kji},1,\Delta_i,\boldsymbol{e}_0)}\cdot\tCF{1}{k}{j}{i}{1}{2})_{ab}&=\frac{1}{\sqrt{2^{r+1}d}}(\Gamma_{12D}C_\Gamma^{-1})_{ab}\bar{I}_{12}^{(d,h_{kji}-1/2,1;\Delta_i)D}\\
&=\frac{1}{\sqrt{2^{r+1}d}}(-2)^{h_{kji}+1/2}(\Delta_i)_{h_{kji}+1/2}(\Delta_i+1-d/2)_{h_{kji}-1/2}\\
&\phantom{=}\qquad\times(h_{kji}+1/2)(\bar{\eta}_3\cdot\Gamma C_\Gamma^{-1})_{ab},\\
\lambda_{\boldsymbol{0}}(\bar{J}_{12;3}^{(d,h_{kji},0,\Delta_i,\boldsymbol{0})}\cdot\tCF{2}{k}{j}{i}{1}{2})_{ab}&=\frac{1}{\sqrt{2^{r+1}}}(C_\Gamma^{-1})_{ab}\bar{I}_{12}^{(d,h_{kji},0;\Delta_i)}\\
&=\frac{1}{\sqrt{2^{r+1}}}(-2)^{h_{kji}}(\Delta_i)_{h_{kji}}(\Delta_i+1-d/2)_{h_{kji}}(C_\Gamma^{-1})_{ab},
}[Eqerer0]
with the help of \eqref{EqJb0ande1}.

Hence, one of the symmetry properties of the tensor structure OPE coefficients corresponds to
\begingroup\makeatletter\def\f@size{10}\check@mathfonts\def\maketag@@@#1{\hbox{\m@th\large\normalfont#1}}%
\eqna{
&\cCF{1}{k}{j}{i}\frac{1}{\sqrt{2^{r+1}d}}2(-2)^{h_{kji}+1/2}(\Delta_i)_{h_{kji}+1/2}(\Delta_i+1-d/2)_{h_{kji}-1/2}(h_{kji}+1/2)(\bar{\eta}_3\cdot\Gamma C_\Gamma^{-1})_{cb}\\
&\qquad+\cCF{2}{k}{j}{i}\frac{1}{\sqrt{2^{r+1}}}(-2)^{h_{kji}}(\Delta_i)_{h_{kji}}(\Delta_i+1-d/2)_{h_{kji}}(\bar{\eta}_3\cdot\Gamma\bar{\eta}_2\cdot\Gamma C_\Gamma^{-1})_{cb}\\
&=\cCF{1}{i}{j}{k}\frac{(-1)^{r(r+1)/2}}{2^{r+1}\sqrt{d}}2(-2)^{h_{ijk}+1/2}(\Delta_k+1/2)_{h_{ijk}-1/2}(\Delta_k+1/2-d/2)_{h_{ijk}+1/2}(h_{ijk}+1/2)(\bar{\eta}_3\cdot\Gamma C_\Gamma^{-1})_{cb}\\
&\phantom{=}\qquad+\cCF{2}{i}{j}{k}\frac{(-1)^{(r+1)(r+2)/2}}{2^{r+1}}(-2)^{h_{ijk}}(\Delta_k+1/2)_{h_{ijk}}(\Delta_k+1/2-d/2)_{h_{ijk}}(\bar{\eta}_3\cdot\Gamma\bar{\eta}_2\cdot\Gamma C_\Gamma^{-1})_{cb},
}
\endgroup
after interchanging $\eta_1\leftrightarrow\eta_3$, rewriting the half-projectors in terms of the new embedding space coordinates, renaming the dummy indices, and manipulating the expressions to put them in the same form.  This result agrees with \eqref{EqPerm} and shows that in this specific case, the tensor structure OPE coefficients of the permuted three-point correlation function are simple rescalings of the corresponding tensor structure OPE coefficients of the original three-point function.


\subsubsection{Scalar-Spinor-\texorpdfstring{$(\boldsymbol{e}_1+\boldsymbol{e}_r)$}{(e1+er)}}

So far, we have computed three-point correlation functions for quasi-primary operators in pure irreducible representations.  That choice was made for simplicity, as the main goal of this paper is to show how the formalism of interest works at the level of three-point correlation functions.  However, as the formalism is completely general, there is absolutely no difference in the treatment of quasi-primary operators in pure and in mixed irreducible representations, as should already be clear from the $\bar{J}$-functions \eqref{EqJbSubs}.

To illustrate this point, we now study a three-point correlation function with an exchanged quasi-primary operator in a mixed irreducible representation, namely $\boldsymbol{e}_1+\boldsymbol{e}_r$.  For simplicity, the computation is again performed in odd spacetime dimensions,\footnote{Moreover, we assume $d>3$ since $\boldsymbol{e}_1+\boldsymbol{e}_r\to2\boldsymbol{e}_1+\boldsymbol{e}_1=3\boldsymbol{e}_1$ in three spacetime dimensions, which is not a mixed irreducible representation.} the even-dimensional case being mostly equivalent.

We consider the $[\boldsymbol{0},\boldsymbol{e}_r,\boldsymbol{e}_1+\boldsymbol{e}_r]$ three-point correlation function.  In this case, the tensor product decomposition contains only two symmetric-traceless irreducible representations
\eqn{\boldsymbol{0}\otimes\boldsymbol{e}_r\otimes(\boldsymbol{e}_1+\boldsymbol{e}_r)\supset2\boldsymbol{e}_1\oplus\boldsymbol{e}_1,}
and therefore only two tensor structures.  These are given by
\eqn{
\begin{gathered}
(\tCF{1}{i}{j}{k}{1}{2})_{bc'C'D_1D_2}=\sqrt{\frac{1}{2^r(d+2)(d-1)}}\left[\A_{12C'(D_1}(\Gamma_{12D_2)}C_\Gamma^{-1})_{bc'}-\frac{1}{d}\A_{12D_1D_2}(\Gamma_{12C'}C_\Gamma^{-1})_{bc'}\right],\\
(\tCF{2}{i}{j}{k}{1}{2})_{bc'C'D}=\sqrt{\frac{d-1}{2^{r+1}d(d-2)}}\left[\A_{12C'D}(C_\Gamma^{-1})_{bc'}-\frac{1}{d-1}(\Gamma_{12C'D}C_\Gamma^{-1})_{bc'}\right].
\end{gathered}
}
It is now a trivial matter to compute the three-point correlation function with the help of \eqref{EqJbe1+er} and $\lambda_{\boldsymbol{e}_1+\boldsymbol{e}_r}=1/\sqrt{2^{r+1}(d-1)}$.  At this point, it is an elementary mathematical exercise to use the contiguous relations \eqref{EqCont} as well as the Clifford algebra to simplify the results, as was done for the previous examples.


\subsubsection{Scalar-\texorpdfstring{$($}{(}Self-dual\texorpdfstring{$)$}{)}-\texorpdfstring{$($}{(}Anti-self-dual\texorpdfstring{$)$}{)} and Permutations in Four Spacetime Dimensions}

As a final example, we describe three-point correlation functions with quasi-primary operators in (anti-)self-dual representations.  The aim of this example is to elucidate the embedding space coordinate ordering in \eqref{EqPTS}.  Indeed, from \eqref{EqTSsub} general (anti-)self-dual representations are the only irreducible representations where the embedding space coordinate ordering is important.

To be specific, we will work in four spacetime dimensions where $\mathscr{K}=\pm i$, with the sign fixed by the definition of the gamma matrices, $\gamma^{\mu_1\cdots\mu_4}=\mathscr{K}\epsilon^{\mu_1\cdots\mu_4}\1$.  Moreover, we will investigate the $[\boldsymbol{0},2\boldsymbol{e}_1,2\boldsymbol{e}_2]$ correlation function with the help of the position space projectors
\eqn{(\hat{\mathcal{P}}^{2\boldsymbol{e}_1})_{\mu_2\mu_1}^{\phantom{\mu_2\mu_1}\nu_1\nu_2}=\frac{1}{2}\delta_{[\mu_1}^{\phantom{[\mu_1}\nu_1}\delta_{\mu_2]}^{\phantom{\mu_2]}\nu_2}+\frac{\mathscr{K}}{4}\epsilon_{\mu_1\mu_2}^{\phantom{\mu_1\mu_2}\nu_2\nu_1},\qquad(\hat{\mathcal{P}}^{2\boldsymbol{e}_2})_{\mu_2\mu_1}^{\phantom{\mu_2\mu_1}\nu_1\nu_2}=\frac{1}{2}\delta_{[\mu_1}^{\phantom{[\mu_1}\nu_1}\delta_{\mu_2]}^{\phantom{\mu_2]}\nu_2}-\frac{\mathscr{K}}{4}\epsilon_{\mu_1\mu_2}^{\phantom{\mu_1\mu_2}\nu_2\nu_1}.}
From the tensor product decomposition,
\eqn{\boldsymbol{0}\otimes2\boldsymbol{e}_1\otimes2\boldsymbol{e}_2=2\boldsymbol{e}_1+2\boldsymbol{e}_2,}
there is only one tensor structure, which is
\eqn{(\tCF{}{i}{j}{k}{1}{2})_{B_2B_1C'_2C'_1}=(\hat{\mathcal{P}}_{21}^{2\boldsymbol{e}_1})_{B_2B_1}^{\phantom{B_2B_1}C''_1C''_2}g_{C''_2C'_2}g_{C''_1C'_1}=\frac{1}{2}\A_{12[B_1C'_1}\A_{12B_2]C'_2}+\frac{\mathscr{K}}{4}\epsilon_{21B_1B_2C'_2C'_1},}[EqTS02e12e2]
following \eqref{EqPTS}, where the $B_1$, $B_2$ indices are anti-symmetrized.  Here, the embedding space coordinate ordering in \eqref{EqPTS} was important to obtain a nonvanishing result for the tensor structure.  It is crucial to observe, however, that the tensor structure \eqref{EqTS02e12e2} does not project onto a two-index symmetric-traceless representation, as expected from the tensor product decomposition, but rather onto a scalar.  This observation will be explained shortly.

To shed more light on the embedding space coordinate ordering, we can also look at the permuted three-point correlation function $[2\boldsymbol{e}_1,\boldsymbol{0},2\boldsymbol{e}_2]$.  Due to the ordering in \eqref{EqPTS}, the tensor structure cannot be given by \eqref{EqTS02e12e2}, as the action of the projection operators annihilates it.  Nevertheless, from the tensor product decomposition, there must be a single possible tensor structure.  The only non-vanishing possibility is constructed from $\A_{12A_1C'_1}\A_{12A_2D_1}\A_{12C'_2D_2}$ and is given by
\eqna{
(\tCF{}{j}{i}{k}{1}{2})_{A_2A_1C'_2C'_1D_2D_1}&=\frac{1}{2}\A_{12[C'_1[A_1}\A_{12A_2](D_1}\A_{12D_2)C'_2]}-\frac{\mathscr{K}}{8}\A_{12[A_1(D_1}\epsilon_{12D_2)A_2]C'_2C'_1}\\
&\phantom{=}+\frac{\mathscr{K}}{8}\A_{12[C'_1(D_1}\epsilon_{12A_1A_2C'_2]D_2)}-\frac{1}{4d}\A_{12D_1D_2}\A_{12[A_1C'_1}\A_{12A_2]C'_2},
}[EqTS2e102e2]
after acting with \eqref{EqPTS}.  Here again, the (anti-)symmetrization of indices occurs only on sets sharing the same names: that is the $A$-indices are anti-symmetrized together, the $C'$-indices are anti-symmetrized together, and the $D$-indices are symmetrized together.  This tensor structure does correspond to a two-index symmetric-traceless representation, as expected from the tensor product decomposition.

In summary, the ordering of the embedding space coordinates in \eqref{EqPTS} is important for general (anti-)self-dual representations since they contain $\epsilon_{12}$, which is the only quantity that is not symmetric under the interchange $\eta_1\leftrightarrow\eta_2$.  In the context of this example, we see that the tensor structure \eqref{EqTS2e102e2} corresponds to the expected tensor product decomposition $\boldsymbol{0}\otimes2\boldsymbol{e}_1\otimes2\boldsymbol{e}_2$ while \eqref{EqTS02e12e2} does not.  Rather, the tensor structure \eqref{EqTS02e12e2} corresponds to the tensor product decomposition $\boldsymbol{0}\otimes2\boldsymbol{e}_2\otimes2\boldsymbol{e}_2\supset\boldsymbol{0}$ due to the antisymmetry of $\epsilon_{21}=-\epsilon_{12}$, as dictated by \eqref{EqPTS}.

It is now a trivial matter both to compute the contribution \eqref{EqThreePt} to the corresponding three-point correlation function and to determine the relation among the permuted coefficients as in \eqref{EqPerm}.


\section{Three-Point Correlation Functions without the OPE}\label{SecThreenoOPE}

The OPE enables the computation of all three-point correlation functions in a straightforward manner.  However, it is clear from the examples in the previous section that the three-point correlation function contributions in \eqref{EqThreePt}, obtained using the orthonormalized OPE tensor structures, are not necessarily the simplest objects.  This complexity arises due to the introduction of a third embedding space coordinate.  In this section, we exploit the knowledge acquired from the OPE to write down the most general three-point correlation functions without directly relying on the OPE.


\subsection{General Three-Point Correlation Functions}

The OPE used on the first two quasi-primary operators leads to three-point correlation functions \eqref{EqCFSub}, repeated here for convenience,
\eqna{
\Vev{\Op{i}{1}\Op{j}{2}\Op{k}{3}}&=\frac{(\mathcal{T}_{12}^{\boldsymbol{N}_i}\Gamma)(\mathcal{T}_{21}^{\boldsymbol{N}_j}\Gamma)(\mathcal{T}_{31}^{\boldsymbol{N}_k}\Gamma)}{\ee{1}{2}{\frac{1}{2}(\tau_i+\tau_j-\chi_k)}\ee{1}{3}{\frac{1}{2}(\chi_i-\chi_j+\tau_k)}\ee{2}{3}{\frac{1}{2}(-\chi_i+\chi_j+\chi_k)}}\\
&\phantom{=}\qquad\cdot\sum_{a=1}^{N_{ijk}}\lambda_{\boldsymbol{N}_k}\cCF{a}{i}{j}{k}\bar{J}_{12;3}^{(d,h_{ijk},n_a,\Delta_k,\boldsymbol{N}_k)}\cdot\tCF{a}{i}{j}{k}{1}{2}.
}
In this result, the three-point correlation function tensor structures are built from $\A_{12}$, $\epsilon_{12}$ and $\Gamma_{12}$, while the $\bar{J}$-function can be expressed in terms of the same quantities along with $\A_{12}\cdot\bar{\eta}_3$ (which plays the role of the OPE differential operator).

Hence, at the level of the three-point correlation functions, the second line can be replaced by a function $F_{ijk}^{12}=F_{ijk}^{12}(\A_{12},\epsilon_{12},\Gamma_{12},\A_{12}\cdot\bar{\eta}_3)$ such that
\eqn{\Vev{\Op{i}{1}\Op{j}{2}\Op{k}{3}}=\frac{(\mathcal{T}_{12}^{\boldsymbol{N}_i}\Gamma)(\mathcal{T}_{21}^{\boldsymbol{N}_j}\Gamma)(\mathcal{T}_{31}^{\boldsymbol{N}_k}\Gamma)}{\ee{1}{2}{\frac{1}{2}(\tau_i+\tau_j-\chi_k)}\ee{1}{3}{\frac{1}{2}(\chi_i-\chi_j+\tau_k)}\ee{2}{3}{\frac{1}{2}(-\chi_i+\chi_j+\chi_k)}}\cdot F_{ijk}^{12}.}[EqCFnoOPE]
The embedding space quantities satisfy the same types of identities as the corresponding position space ones, \textit{e.g.} illustrated by \eqref{EqNTS}.  Thus, one simply needs to find all independent (on the light cone) tensor structures built from $\A_{12}$, $\epsilon_{12}$, $\Gamma_{12}$, and $\A_{12}\cdot\bar{\eta}_3$ which contract all the dummy indices present in $(\mathcal{T}_{12}^{\boldsymbol{N}_i}\Gamma)(\mathcal{T}_{21}^{\boldsymbol{N}_j}\Gamma)(\mathcal{T}_{31}^{\boldsymbol{N}_k}\Gamma)$ in order to generate the most general three-point correlation functions \eqref{EqCFnoOPE}.  Obviously, these tensor structures will be linear combinations of the original tensor structures that naturally arise from the OPE, implying that their coefficients will be linear combinations of the OPE coefficients.

This last observation leads to a straightforward connection to the literature (see \textit{e.g.} \cite{Costa:2011mg}).  Moreover, it provides a recipe that generalizes to quasi-primary operators in arbitrary irreducible representations.  Indeed, from \eqref{EqCFnoOPE} it is possible to contract the embedding space spinor indices with
\eqn{\ee{1}{2}{-(S_k-\xi_k)/2}\ee{1}{3}{[S_i-\xi_i-(S_j-\xi_j)]/2}\ee{2}{3}{[-(S_i-\xi_i)+S_j-\xi_j+S_k-\xi_k]/2}(\mathcal{T}_{21\boldsymbol{N}_i}\Gamma)(\mathcal{T}_{12\boldsymbol{N}_j}\Gamma)(\mathcal{T}_{13\boldsymbol{N}_k}\Gamma),}
resulting in
\eqna{
&\Vev{\Op{i\{aA\}}{1}\Op{j\{bB\}}{2}\Op{k\{cC\}}{3}}\\
&\qquad=\frac{(\bar{\eta}_1\cdot\Gamma\,\hat{\mathcal{P}}_{12}^{\boldsymbol{N}_i}\,\bar{\eta}_2\cdot\Gamma)_{\{aA\}}^{\phantom{\{aA\}}\{A'a'\}}(\bar{\eta}_2\cdot\Gamma\,\hat{\mathcal{P}}_{21}^{\boldsymbol{N}_j}\,\bar{\eta}_1\cdot\Gamma)_{\{bB\}}^{\phantom{\{bB\}}\{B'b'\}}(\bar{\eta}_3\cdot\Gamma\,\hat{\mathcal{P}}_{31}^{\boldsymbol{N}_k}\,\bar{\eta}_1\cdot\Gamma)_{\{cC\}}^{\phantom{\{cC\}}\{C'c'\}}}{\ee{1}{2}{\frac{1}{2}(\tau_i+\tau_j-\tau_k)}\ee{1}{3}{\frac{1}{2}(\tau_i-\tau_j+\tau_k)}\ee{2}{3}{\frac{1}{2}(-\tau_i+\tau_j+\tau_k)}}\\
&\qquad\phantom{=}\times(F_{ijk}^{12})_{\{a'A'\}\{b'B'\}\{c'C'\}},
}[EqCFnoOPEVector]
where it is understood now that the usual embedding space spinor indices of the quasi-primary operators have been replaced by embedding space vector indices (the former dummy indices).\footnote{This contraction could be done at the level of the OPE, but it would be somewhat cavalier since quasi-primary operators would then depend on two embedding space coordinates.}

With $F_{ijk}^{12}=F_{ijk}^{12}(\A_{12},\epsilon_{12},\Gamma_{12},\A_{12}\cdot\bar{\eta}_3)$, \eqref{EqCFnoOPEVector} is the appropriate generalization of the form of three-point correlation functions of quasi-primary operators in arbitrary irreducible representations.  By construction, it is transverse with the proper degree of homogeneity in all three embedding space coordinates, and it naturally projects onto the appropriate irreducible representation for all three quasi-primary operators.

For example, for scalar-scalar-(symmetric-traceless) three-point correlation functions, \eqref{EqCFnoOPEVector} [or \eqref{EqCFnoOPE}] dictates that
\eqn{(F_{ijk}^{12})_{\{C'\}}=(\A_{12}\cdot\bar{\eta}_3)_{C'_\ell}\cdots(\A_{12}\cdot\bar{\eta}_3)_{C'_1},}
which implies
\eqna{
\Vev{\Op{i}{1}\Op{j}{2}\Op{k\{C\}}{3}}&=\frac{c_{ijk}(\hat{\mathcal{P}}_{31}^{\ell\boldsymbol{e}_1})_{\{C\}}^{\phantom{\{C\}}\{C'\}}(\A_{12}\cdot\bar{\eta}_3)_{C'_\ell}\cdots(\A_{12}\cdot\bar{\eta}_3)_{C'_1}}{\ee{1}{2}{\frac{1}{2}(\Delta_i+\Delta_j-\Delta_k+\ell)}\ee{1}{3}{\frac{1}{2}(\Delta_i-\Delta_j+\Delta_k-\ell)}\ee{2}{3}{\frac{1}{2}(-\Delta_i+\Delta_j+\Delta_k-\ell)}}\\
&=\frac{c_{ijk}(\hat{\mathcal{P}}_{31}^{\ell\boldsymbol{e}_1})_{\{C\}}^{\phantom{\{C\}}\{C'\}}(-\bar{\eta}_2)_{C'_\ell}\cdots(-\bar{\eta}_2)_{C'_1}}{\ee{1}{2}{\frac{1}{2}(\Delta_i+\Delta_j-\Delta_k+\ell)}\ee{1}{3}{\frac{1}{2}(\Delta_i-\Delta_j+\Delta_k-\ell)}\ee{2}{3}{\frac{1}{2}(-\Delta_i+\Delta_j+\Delta_k-\ell)}},
}
in agreement with \eqref{EqThreePt00le1}.  Contracting with the usual auxiliary polarization vector $\zeta_3$, which satisfies $\zeta_3^2=\zeta_3\cdot\eta_3=0$, the three-point correlation functions become
\eqna{
\Vev{\Op{i}{1}\Op{j}{2}\zeta_3\cdot\Op{k}{3}}&=\frac{c_{ijk}(-\zeta_3\cdot\A_{13}\cdot\bar{\eta}_2)^\ell}{\ee{1}{2}{\frac{1}{2}(\Delta_i+\Delta_j-\Delta_k+\ell)}\ee{1}{3}{\frac{1}{2}(\Delta_i-\Delta_j+\Delta_k-\ell)}\ee{2}{3}{\frac{1}{2}(-\Delta_i+\Delta_j+\Delta_k-\ell)}}\\
&=\frac{c_{ijk}(\zeta_3\cdot\bar{\eta}_1-\zeta_3\cdot\bar{\eta}_2)^\ell}{\ee{1}{2}{\frac{1}{2}(\Delta_i+\Delta_j-\Delta_k+\ell)}\ee{1}{3}{\frac{1}{2}(\Delta_i-\Delta_j+\Delta_k-\ell)}\ee{2}{3}{\frac{1}{2}(-\Delta_i+\Delta_j+\Delta_k-\ell)}},
}
which is the standard result found in the literature (see \textit{e.g.} \cite{Costa:2011mg}), taking into account the different homogeneity properties of quasi-primary operators in symmetric-traceless representations.

Hence, the application of the OPE leads to all three-point correlation functions, while at the same time explaining how to generalize the usual construction of three-point correlation functions in embedding space to quasi-primary operators in arbitrary irreducible representations.  However, the great power of the OPE in fact lies in its ability to express $M$-point correlation functions in terms of $(M-1)$-point correlation functions, furnishing the complete determination of four-point correlation functions, as will be discussed in a forthcoming publication.


\section{Conclusions}\label{SecConc}

In this paper, we have shown how to construct the three-point function of quasi-primary operators in arbitrary Lorenz representations.  The general result in \eqref{EqCFSub} is obtained directly from the OPE and the corresponding two-point function in \cite{Fortin:2019xyr}.  This general form has several prominent features.  First, the scaling is set to zero in every coordinate, which fixes the powers of the dot products of coordinates.  Second, there are the half-projectors which correspond to each operator and were discussed in detail in \cite{Fortin:2019dnq,Fortin:2019xyr}.  Third, it contains the $\bar{J}$-function, which is obtained by a straightforward substitution shown in \eqref{EqJbSub}.  The $\bar{J}$-function is derived from the action of the differential operator, appearing on the right-hand side of the OPE, on the two-point function of the exchanged operator.  An analogous substitution will be relevant for the computation of four-point functions in a forthcoming publication.  As far as its symmetry properties are concerned, the $\bar{J}$-function is a tensor product of a symmetric-traceless tensor from the derivative operator and Lorentz representations of the exchange operator and its conjugate, and it is homogenous of degree zero in every coordinate.  Fourth, the form contains the tensor structure $\tCF{a}{i}{j}{k}{}{}$ that is present to combine the representations of operators with the derivatives into a Lorentz singlet.  In practice, $\tCF{a}{i}{j}{k}{}{}$ forms a singlet from the dummy indices on the half-projectors with the $\bar{J}$-function.

It is worth stressing the central role of the hatted projection operators $\hat{\mathcal{P}}^{\boldsymbol{N}}_{ij}$ for our construction of these objects.  The half-projectors that relate spinor indices to a vector-spinor index combination can be obtained trivially from the $\hat{\mathcal{P}}$'s.  Further, the $\bar{J}$-function is computed from the product of $\hat{\mathcal{P}}^{\boldsymbol{N}}$'s at different coordinate points \eqref{EqJbSub}.  Finally, the tensor structure obeys \eqref{EqPTS} because it intertwines four different representations.  Constructing the hatted projectors is not straightforward, but can be done iteratively, as discussed \textit{e.g.} in \cite{Fortin:2019dnq}.  One can imagine that the entire process of constructing the hatted projectors, implementing the conformal substitutions, and contracting indices into Lorentz singlets could be handled by a suitable computer program.

We have shown several detailed examples of three-point functions for different choices of operators and have explicitly illustrated how to find all the necessary ingredients.  Although most of the calculations in this work have been performed in embedding space, in a few cases, we have verified that the projections to position space reproduce known results, which provides a consistency check on our approach.  Moreover, since the OPE in embedding space is inherently not symmetric, we have also studied how three-point functions behave under permutations of operators, which led to relations among the OPE coefficients under index permutations \eqref{EqPerm}.  Clearly, using the OPE is not absolutely necessary for finding three-point functions.  A general formula bypassing the OPE is presented in \eqref{EqCFnoOPEVector}, which again highlights the importance of the hatted projection operators.

Following the study of two-point functions in \cite{Fortin:2019xyr}, this work constitutes the next step in the program of applying the embedding space OPE formalism developed in \cite{Fortin:2019fvx,Fortin:2019dnq} to systematically compute correlation functions.  While the OPE is not vital for constructing two- and three-point functions, many results obtained here will be useful for the much more important case of four-point functions and, looking ahead, to applications of these results.


\ack{
The authors would like to thank Wenjie Ma for useful discussions.  The work of JFF and VP is supported by NSERC and FRQNT.  WS would like to thank the KITP Santa Barbara for its hospitality, his work was supported in part by the National Science Foundation under Grant No. NSF PHY-1748958.
}


\bibliography{ThreePtFcts}

\end{document}